\PassOptionsToPackage{numbers,super,sort&compress}{natbib}
\documentclass[pdflatex]{sn-jnl}
\usepackage{natbib}
\usepackage{lmodern}
\usepackage{one2n-article}
\usepackage{multirow}
\usepackage{textcomp}
\usepackage{placeins}
\usepackage{titletoc}
\usepackage{chapterbib}
\usepackage{booktabs}
\usepackage{hyperref}
\makeatletter
\long\def\@makecaption#1#2{%
  \vskip6pt
  \begingroup\small\raggedright
  {\bfseries #1.} #2\par
  \endgroup\vskip6pt}
\makeatother
\renewcommand{\figurename}{Fig.}
\renewcommand{\tablename}{Table}
\setcitestyle{numbers,super,open={},close={}}
\onearticlerunningtitle{One2N: Recursive refinement breaks the resolution barrier for Earth-system AI}

\ifdefined\OneTwoNMainOnly
\makeatletter
\newcounter{ttl@toc@supp}
\makeatother
\fi
\ifdefined\OneTwoNSuppOnly
\makeatletter
\newcounter{ttl@toc@supp}
\makeatother
\fi
\begin{document}
\raggedbottom

\title[One2N recursive refinement]{One2N: Recursive refinement breaks the resolution barrier for Earth-system AI}

\author[1,2]{\fnm{Yanyan} \sur{Tan}}\email{tanyy@itpcas.ac.cn}
\equalcont{These authors contributed equally to this work.}
\author*[1,2]{\fnm{Xiangjun} \sur{Tian}}\email{Tianxj@itpcas.ac.cn}
\equalcont{These authors contributed equally to this work.}
\author*[3]{\fnm{Can} \sur{Wang}}\email{wangcan@bistu.edu.cn}
\author[4]{\fnm{Hongqin} \sur{Zhang}}\email{hqzhang1112@mail.iap.ac.cn}
\author[4]{\fnm{Yinhai} \sur{Luo}}\email{luoyinhai20@mails.ucas.ac.cn}
\author[1]{\fnm{Min} \sur{Zhao}}\email{zhaomin@itpcas.ac.cn}
\author[7]{\fnm{Zhe} \sur{Jin}}\email{zhejin@pku.edu.cn}
\author[1]{\fnm{Xingchao} \sur{Yang}}\email{yangxc@itpcas.ac.cn}
\author[1]{\fnm{Zaifei} \sur{Wang}}\email{wangzaifei@itpcas.ac.cn}
\author[5]{\fnm{Banglin} \sur{Zhang}}\email{zhangbanglin24@nudt.edu.cn}
\author[3]{\fnm{Aiwen} \sur{Wang}}\email{wangaiwen@bistu.edu.cn}
\author[6]{\fnm{Jing} \sur{Zhou}}\email{zoujing@qlu.edu.cn}
\author*[1]{\fnm{Tao} \sur{Wang}}\email{Twang@itpcas.ac.cn}
\author[1]{\fnm{Baosheng} \sur{An}}\email{anbaosheng@itpcas.ac.cn}
\author[1,2]{\fnm{Xuliang} \sur{Zhuang}}\email{xlzhuang@itpcas.ac.cn}
\author[7]{\fnm{Shilong} \sur{Piao}}\email{slpiao@pku.edu.cn}
\author[1]{\fnm{Tandong} \sur{Yao}}\email{tdyao@itpcas.ac.cn}

\affil*[1]{\orgdiv{State Key Laboratory of Tibetan Plateau Earth System, Environment and Resources (TPESER)}, \orgname{Institute of Tibetan Plateau Research, Chinese Academy of Sciences}, \orgaddress{\city{Beijing}, \postcode{100101}, \country{China}}}
\affil[2]{\orgname{University of Chinese Academy of Sciences}, \orgaddress{\city{Beijing}, \postcode{101408}, \country{China}}}
\affil[3]{\orgdiv{College of Mathematical Science}, \orgname{Beijing Information Science and Technology University}, \orgaddress{\city{Beijing}, \postcode{100192}, \country{China}}}
\affil[4]{\orgname{Institute of Atmospheric Physics, Chinese Academy of Sciences}, \orgaddress{\city{Beijing}, \postcode{100029}, \country{China}}}
\affil[5]{\orgdiv{College of Meteorology and Oceanography}, \orgname{National University of Defense Technology}, \orgaddress{\city{Changsha}, \postcode{410073}, \country{China}}}
\affil[6]{\orgdiv{State Key Laboratory of Physical Oceanography, Institute of Oceanographic Instrumentation}, \orgname{Qilu University of Technology (Shandong Academy of Sciences)}, \orgaddress{\city{Qingdao}, \postcode{266100}, \country{China}}}
\affil[7]{\orgdiv{Institute of Carbon Neutrality, College of Urban and Environmental Sciences}, \orgname{Peking University}, \orgaddress{\city{Beijing}, \postcode{100871}, \country{China}}}

\abstract{
Earth-system AI models approach kilometre-scale resolution, but halving grid spacing quadruples the grid-point count and end-to-end training cost. We introduce One2N, a recursive incremental framework extending a coarse model through \(N\) levels of cross-resolution residual correction. At each level the parent is frozen and excluded from backpropagation, so its activations and optimizer states are never retained; only a tier-specific module is trained. A single refinement level on the NeuralGCM-0.7\textdegree{} baseline yields the two-level One2N-0.07\textdegree{}, trained on one 80 GB GPU. Against HRES-0.07\textdegree{} and ERA5-0.25\textdegree{} analyses it matches its parent’s tropospheric skill while reducing upper-stratospheric errors and resolving finer structure. Tropical-cyclone track errors nearly match the parent’s and are lower than Pangu-Weather-0.25\textdegree{} at medium range; RMSE is lower than that model’s on 93.46\% of common targets. One2N thus turns an existing forecasting system, or an inexpensively trained coarse baseline, into a reusable foundation for high-resolution development, breaking the resolution barrier.
}

\maketitle
\pagestyle{fancy}
\thispagestyle{titlepage}
\hypersetup{hidelinks}

Earth system models are approaching a resolution barrier. As horizontal grid spacing reaches kilometre scales, processes such as convective storms and orographic precipitation can be represented more explicitly, but rising computational and storage costs constrain routine global kilometre-scale modelling\cite{prein2015review,satoh2019global,schulthess2019reflecting,wedi2020baseline}. Global kilometre-scale simulations offer the prospect of better representing extreme events and providing more spatially detailed forecast information for societal decision-making. Realizing these benefits requires computationally affordable pathways to finer resolutions\cite{bauer2015quiet,stevens2019dyamond,schar2020kilometer,dueben2018challenges}.

Advances in physics-based and data-driven modelling are extending this frontier. MPAS has demonstrated parallel scaling on a global 3 km mesh, whereas ICON has achieved coupled global Earth system simulations at 1.25 km grid spacing\cite{heinzeller2016towards,klocke2025computing}. Data-driven systems, including Pangu-Weather, GraphCast and FuXi, have matched or exceeded the skill of ECMWF’s operational forecasts across multiple medium-range forecasting tasks in their respective evaluations, while reducing inference times to seconds or minutes\cite{bi2023accurate,lam2023learning,chen2023fuxi}. Aurora and FengWu-GHR have further extended global forecasting to 0.1\textdegree{} and 0.09\textdegree{} grids, respectively, through adaptation of pretrained models, demonstrating the value of existing predictive knowledge for high-resolution modelling\cite{bodnar2025foundation,han2024fengwu}.

Fast inference, however, does not remove the resource demands of high-resolution training. Finer grids generally increase the sizes of data arrays and intermediate activations, while joint optimization across components or resolution levels may add activation-storage or recomputation costs\cite{bodnar2025foundation,chen2016sublinear}. Hybrid modelling offers another route to reusing established dynamical structures\cite{wattmeyer2021correcting,kochkov2021accelerated,pathak2020augment,farchi2021correcting}: NeuralGCM-0.7\textdegree{} couples a dynamical core with learned physical parameterizations, whereas HMgNO has demonstrated a state-correction framework on a 1.40625\textdegree{} grid without requiring gradients to propagate through the dynamical core\cite{kochkov2024neural,hu2026hybrid}. 

These advances establish the value of transferring pretrained knowledge and reusing dynamical components. For established forecasting systems, a further question is how to extend their existing skill to finer-grid global autoregressive forecasting while keeping the parent predictor’s complete mapping fixed and avoiding differentiation through its internal computations. This extension requires reconciling state representations across grids and addressing the propagation of correction errors during rollout\cite{brenowitz2018prognostic,rasp2020coupled}. Finer spatial detail also complicates verification: grid-point RMSE can favour smoother forecasts when spatially displaced features incur a double penalty\cite{gilleland2009intercomparison, pmlr-v267-subich25a}. Resolution refinement should therefore be assessed through forecast errors and spatial structure on the finer grid, together with the retention of the parent model’s skill during autoregressive rollout. Tropical cyclone track prediction provides a complementary assessment of forecast performance for high-impact weather.

\begin{figure*}[htbp] 
\centering 
\includegraphics[width=\linewidth,height=0.73\textheight,keepaspectratio]{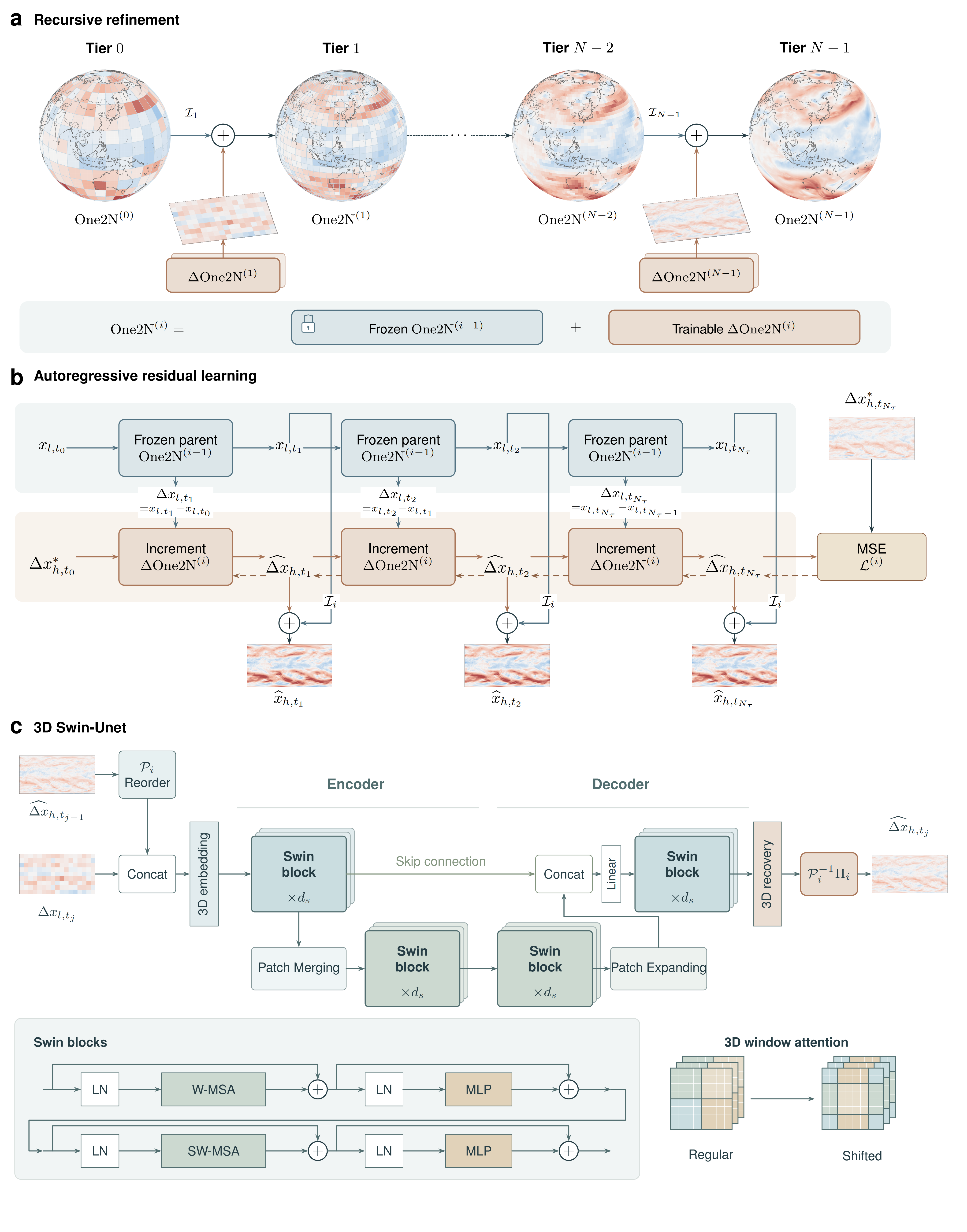} 
\caption[Recursive refinement and residual learning in One2N]{\textbf{Recursive refinement and residual learning in One2N.}
\textbf{a)} At each tier, the predictor from the preceding tier is frozen. Its forecast is mapped to a finer grid using \(\mathcal I_i\) and combined with a learned residual correction. This study evaluates a single refinement level (\(N=2\)), corresponding to a two-level model comprising the frozen parent and one residual correction tier. Tier 0 is the single baseline, obtained independently at coarse resolution; each later tier is trained on top of its own frozen parent.
\textbf{b)} The recurrent correction module uses the parent’s temporal increment \(\Delta x_{l,t_j}\) and the preceding high-resolution residual to predict the next residual. Reference residuals are defined as \(\Delta x^*_{h,t_j}=x^*_{h,t_j}-\mathcal I_i(x_{l,t_j})\), where \(\mathcal I_i\) maps the parent state to the finer grid. The rollout starts from \(\Delta x^*_{h,t_0}\) and uses predicted residuals \(\widehat{\Delta x}_{h,t_j}\) thereafter. The residual may represent both spatial refinement and correction of parent-model errors. At each step, adding the predicted residual to the mapped parent forecast reconstructs the high-resolution state. The correction module is unrolled for \(N_\tau=3\) steps of 2~h, with a latitude-weighted mean squared error (MSE) loss evaluated only at the final step between \(\widehat{\Delta x}_{h,t_{N_\tau}}\) and \(\Delta x^*_{h,t_{N_\tau}}\). Subscripts \(l\) and \(h\) denote the parent and finer grids, respectively; hats denote predictions and asterisks denote reference quantities.
\textbf{c)} The 3D Swin-Unet correction module comprises spatial rearrangement, concatenation, and an encoder–decoder architecture with repeated Swin blocks and one encoder--decoder skip connection. Here, \(\mathcal P_i\) rearranges spatial blocks into channels, \(\Pi_i\) selects residual channels before inverse rearrangement, and \(d_s\) denotes the number of blocks at stage \(s\). The insets show window-based and shifted-window multi-head self-attention (W-MSA and SW-MSA), layer normalization (LN), multilayer perceptrons (MLPs) and residual connections. Fields and architectural dimensions are schematic. Implementation details are provided in Supplementary Section~\ref{sec:supp_note3}.}
\label{fig:one2n_framework} 
\end{figure*}

Here we introduce One2N, a recursive incremental training framework that organizes resolution refinement into \(N\) successive levels (levels 0 to \(N-1\); Fig.~\ref{fig:one2n_framework}a). At each level, the complete predictor from the preceding level serves as a fixed parent model; only a separately trained residual module is optimized, with all preceding levels excluded from backpropagation (Fig.~\ref{fig:one2n_framework}b). The correction module is unrolled along the frozen parent trajectory, using the parent’s temporal increments and the preceding high-resolution residual to predict the next residual, which is added to the mapped parent forecast to reconstruct the state on the target grid. Compared with joint training of the full model hierarchy, this design avoids retaining the frozen parent model’s internal activations for backpropagation or storing its parameter gradients and optimizer states. It therefore reduces backpropagation-related memory requirements and prevents this component of storage from accumulating across frozen levels. Parent and child models interact through state interfaces, without modifying the parent’s internal architecture or requiring it to be differentiable. 

Throughout, tier 0 denotes the single baseline of a One2N hierarchy, and the parent of a tier is the frozen model immediately above it. A hierarchy therefore has one baseline but one parent per refinement tier; in the two-level demonstration below, the baseline is also the parent of the only refinement tier. Two starting points are therefore available, and only one baseline is required. It can be an existing forecasting system, reused unchanged; or a coarse-resolution model trained from scratch, whose cost is set by the coarse grid rather than by the target grid. In both cases the incremental strategy carries the remaining learning, so that kilometre-scale development becomes a matter of adding a tier rather than of rebuilding a system.

We demonstrate One2N through a single refinement level, yielding a two-level model, using NeuralGCM-0.7\textdegree{}\cite{kochkov2024neural} as the coarse baseline (tier 0), and hence as the parent of the single refinement tier, together with a residual module based on Swin-Unet\cite{cao2023swinunet,liu2021swin} (Fig.~\ref{fig:one2n_framework}c). The resulting One2N-0.07\textdegree{} model predicts five atmospheric variables on 25 pressure levels with a 2~h update interval on a global grid nominally spaced at 0.07\textdegree{}. The correction module is trained on HRES-0.07\textdegree{} analyses\cite{ecmwf2016hresdata}, with training completed in approximately four weeks on a single NVIDIA A800 GPU with 80 GB of memory, including online integration of the frozen parent. A comparison of reported training hardware and training times with other forecasting models is provided in Supplementary Section~\ref{subsec:supp_training_resources}. Global forecast skill is evaluated primarily against HRES-0.07\textdegree{} analyses\cite{ecmwf2016hresdata}, with ERA5-0.25\textdegree{}\cite{hersbach2020era5} used to assess robustness to the verification reference; tropical cyclone track forecasts provide a complementary assessment. The results show broadly comparable skill to the parent on representative tropospheric diagnostics, alongside reduced errors in selected upper-stratospheric variables. These results provide empirical support for an incremental formulation of global resolution refinement, in which residual evolution is learned along a fixed parent trajectory, enabling established forecasting systems to serve as reusable foundations for higher-resolution model development.

\section{Results}\label{sec3}

\subsection{Forecast verification results}\label{subsec:forecast_result}

We evaluated global 10-day forecasts from 120 matched initializations in the independent 2023 test set using latitude-weighted root-mean-square error (RMSE) and anomaly correlation coefficient (ACC)\cite{rasp2024weatherbench2}. Comparisons with the parent model, NeuralGCM-0.7\textdegree{}\cite{kochkov2024neural}, which also serves as the parent of the refinement tier, covered five atmospheric variables at 25 pressure levels, whereas comparisons with Pangu-Weather-0.25\textdegree{}\cite{bi2023accurate} were restricted to five variables at 13 common pressure levels. Pangu-Weather-0.25\textdegree{} was evaluated by repeatedly applying its 3~h model, while One2N-0.07\textdegree{} and NeuralGCM-0.7\textdegree{} used 2~h updates; forecasts were sampled at common verification lead times. For atmospheric variables shared by all three models, initial conditions were derived from HRES-0.07\textdegree{} analyses at the same initialization time. Additional model-specific inputs and their processing were described in Supplementary Section~\ref{subsec:forecast_initialization}. All models were compared on common verification grids: 0.07\textdegree{} for HRES-0.07\textdegree{}-based scores and 0.25\textdegree{} for ERA5-0.25\textdegree{}-based scores. Details of the data splits, model configurations, remapping procedures and metric calculations are provided in Supplementary Sections~\ref{sec:supp_note1},~\ref{sec:supp_note2} and~\ref{sec:supp_note5}.

In this comparison, One2N-0.07\textdegree{} achieved lower RMSE than the evaluated Pangu-Weather-0.25\textdegree{} configuration for most common verification targets. Of the 1,300 targets (five variables, 13 common pressure levels and 20 non-zero forecast lead times), 1,215 (93.46\%) had lower RMSE. For Z500, T850 and Q700, the mean lead-time-specific relative RMSE reductions over 12--240 h were 27.88\%, 22.90\% and 18.46\%, respectively (Fig.~\ref{fig:forecast_skill}A). The full comparison across the 13 common pressure levels is shown in Supplementary Fig.~\ref{sfig:pangu_hres_full}. 

\begin{figure*}[htbp]
\centering
\includegraphics[width=\linewidth,height=0.80\textheight,keepaspectratio]{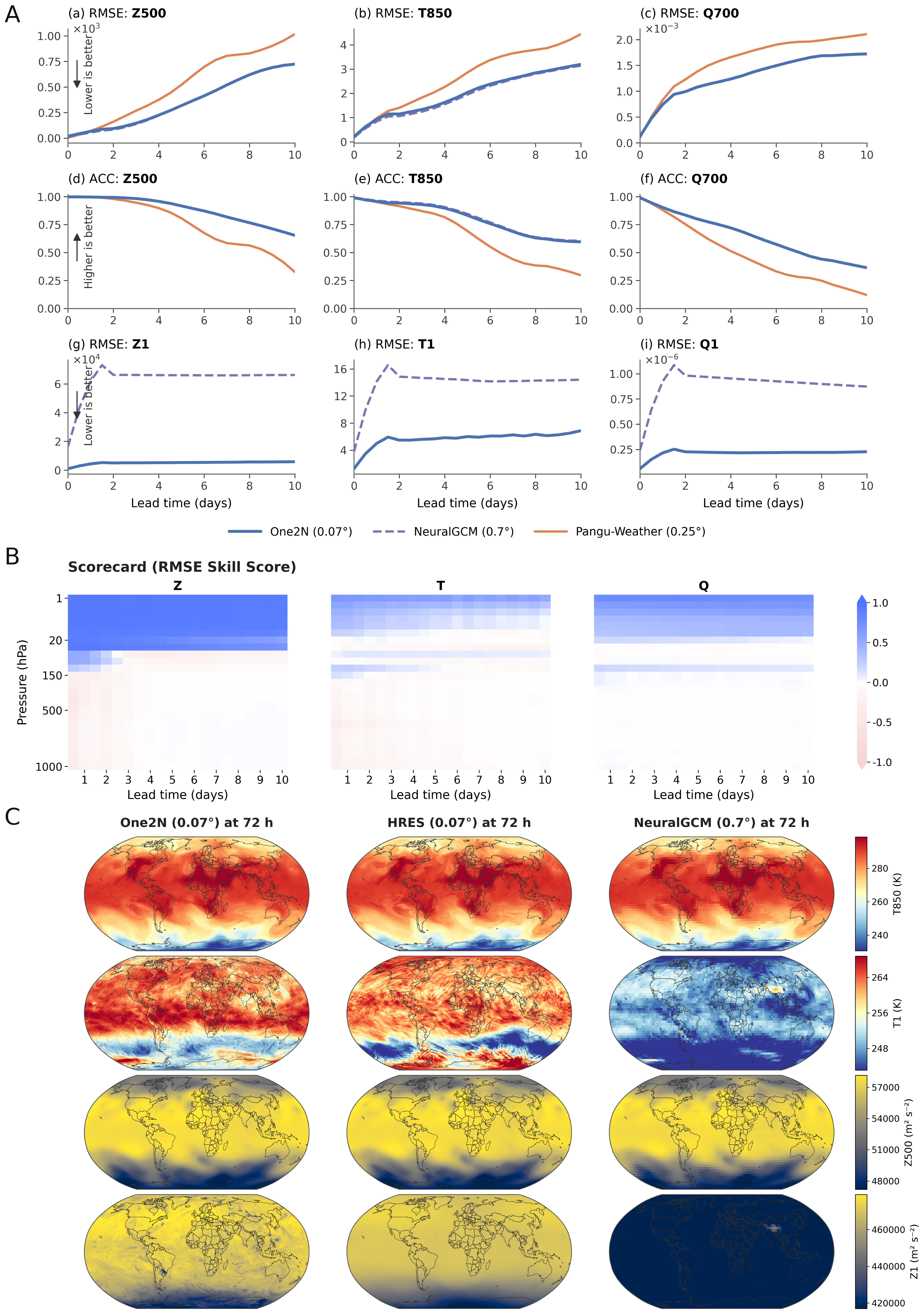}
\caption[Global forecast skill and comparison of 72-h forecast fields]{\textbf{Global forecast skill and comparison of 72-h forecast fields.}
\textbf{(A)} Global forecast skill of One2N-0.07\textdegree{} (solid blue), NeuralGCM-0.7\textdegree{} (dashed purple) and Pangu-Weather-0.25\textdegree{} (solid orange), evaluated against HRES-0.07\textdegree{} analyses. The top and middle rows show RMSE and ACC, respectively, for Z500, T850 and Q700; the bottom row shows RMSE for Z1, T1 and Q1. Z, T and Q denote geopotential, temperature and specific humidity, respectively, and the numerical suffixes indicate pressure levels in hPa. The bottom row includes only One2N-0.07\textdegree{} and NeuralGCM-0.7\textdegree{}, as Pangu-Weather-0.25\textdegree{} does not provide forecasts at 1 hPa. Lower RMSE and higher ACC indicate greater forecast skill. RMSE skill scores for all five variables across the 25 pressure levels are shown in Supplementary Fig.~\ref{sfig:vertical_scorecard}.
\textbf{(B)} RMSE skill scores of One2N-0.07\textdegree{} relative to NeuralGCM-0.7\textdegree{} for Z, T and Q as a function of pressure level and forecast lead time. The skill score is defined as \(1-\mathrm{RMSE}_{\text{One2N-0.07\textdegree{}}}/\mathrm{RMSE}_
{\text{NeuralGCM-0.7\textdegree{}}}\). Positive values (blue) and negative values (red) indicate lower and higher RMSE for One2N-0.07\textdegree{}, respectively; zero (white) indicates equal RMSE. Results for additional tropospheric variables are shown in Supplementary Fig.~\ref{sfig:vertical_scorecard}.
\textbf{(C)} Global 72-h forecast fields from a common initialization and reference analyses at the corresponding valid time. Columns show, from left to right, One2N-0.07\textdegree{} forecasts, HRES-0.07\textdegree{} analyses and NeuralGCM-0.7\textdegree{} forecasts. Rows show, from top to bottom, T850, T1, Z500 and Z1. Each row uses a shared colour scale, with variables and units indicated on the right. }
\label{fig:forecast_skill}
\end{figure*}

Relative to its parent model NeuralGCM-0.7\textdegree{}, the improvements from One2N-0.07\textdegree{} exhibited a clear vertical structure. Forecast skill for representative tropospheric metrics, including Z500, T850 and Q700, remained broadly comparable to that of the parent model, whereas the largest error reductions occurred in upper-stratospheric geopotential and temperature  (Fig.~\ref{fig:forecast_skill}A,B; Supplementary Fig.~\ref{sfig:vertical_scorecard}). Across the evaluated pressure levels between 1 and 7 hPa, the lead-time-averaged relative RMSE reductions for geopotential ranged from 91.10\% to 92.12\%; the corresponding reduction for temperature at 1 hPa was 59.14\%. These reductions in geopotential and temperature at 1 hPa were evident at 12 h and persisted through 240 h. Evaluation against ERA5-0.25\textdegree{} also supported the advantages over Pangu-Weather-0.25\textdegree{} for most common targets and the marked reductions in selected upper-level errors relative to the parent model (Supplementary Figs.~\ref{sfig:vertical_scorecard_neuralgcm_0.25} and~\ref{sfig:vertical_scorecard_pangu_0.25}).

To investigate this vertical contrast, we examined forecast errors against HRES-0.07\textdegree{} and ERA5-0.25\textdegree{} analyses, alongside differences between the two analyses, on a common 0.25\textdegree{} grid (Supplementary Section~\ref{sec:one2n_mechanism}). For the representative tropospheric metrics examined, the parent model's mean bias was small relative to its RMSE, and One2N-0.07\textdegree{} retained broadly comparable forecast skill. By contrast, the parent forecasts of geopotential and temperature at 1--7 hPa exhibited large mean biases that, for each variable, had the same sign against both analyses, with magnitudes substantially exceeding the corresponding analysis-to-analysis differences. These reference-consistent biases identify a systematic component of the parent model's upper-level error that provides a clear target for residual correction. Consistent with this interpretation, One2N-0.07\textdegree{} reduced upper-level RMSE against both references, with corresponding reductions in normalized RMSE in the pressure-band comparisons (Supplementary Figs.~\ref{fig:supp_vertical_diagnostics}--\ref{fig:supp_pressure_rmse}). The alignment between the parent model’s error structure and the largest forecast gains supports a contribution from systematic bias correction. These findings support cross-resolution residual learning as a means of extending a mature forecasting model to finer resolutions while retaining its predictive strengths and addressing its existing systematic errors.

To illustrate the spatial characteristics of the forecasts, we further compared deterministic 72~h forecast fields with the corresponding HRES-0.07\textdegree{} analyses (Fig.~\ref{fig:forecast_skill}C). One2N-0.07\textdegree{} retained the broad tropospheric patterns represented by NeuralGCM-0.7\textdegree{} while exhibiting finer spatial variability. At 1 hPa, its geopotential and temperature fields were also substantially closer to the reference in their background values, consistent with the large reductions in statistical error. These comparisons illustrate the coexistence of large-scale pattern retention, finer spatial variability and improved upper-level states in the refined forecasts.

\subsection{Tropical cyclone track forecast evaluation}  
To assess forecast performance in a high-impact weather setting, we evaluated tropical cyclone track forecasts from One2N-0.07\textdegree{}, Pangu-Weather-0.25\textdegree{} and NeuralGCM-0.7\textdegree{} against best-track positions from version v04r01 of the International Best Track Archive for Climate Stewardship (IBTrACS)\cite{knapp2010ibtracs,gahtan2024ibtracsdata}. This assessment included case studies of Haikui, Saola and Bolaven and an aggregate evaluation of 185 cyclone--initialization pairs from 84 tropical cyclones in 2023. Track error was defined as the great-circle distance between the forecast cyclone centre and the best-track position at the same valid time. Details of cyclone tracking and sample selection are provided in Supplementary Sections~\ref{subsec:tc_tracking} and~\ref{subsec:tc_evaluation_protocol}.

\begin{figure*}[htbp]
\centering
\includegraphics[width=\linewidth,height=0.80\textheight,keepaspectratio]{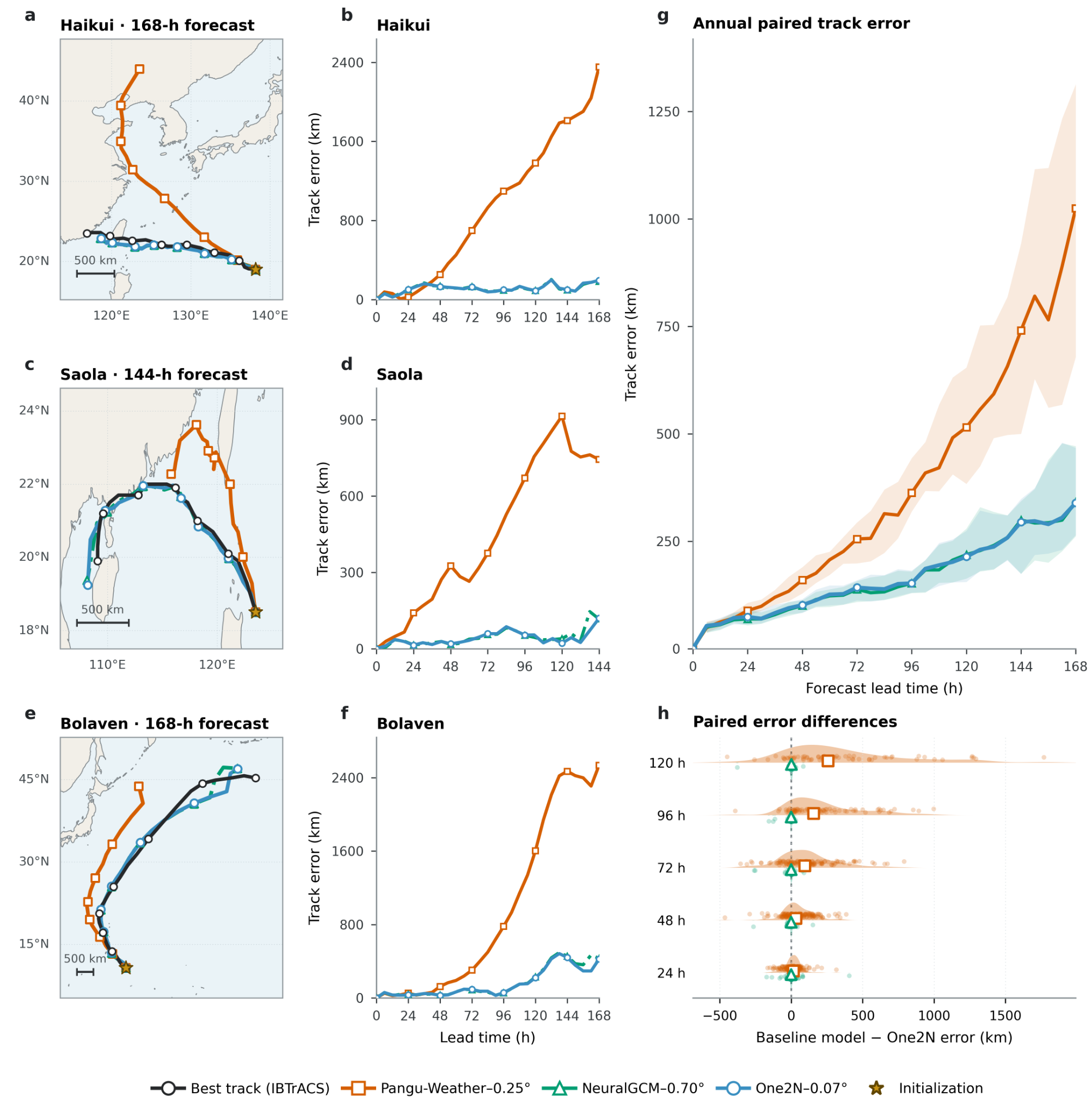}
\caption[Case studies and statistical evaluation of tropical cyclone track forecasts]{\textbf{Case studies and statistical evaluation of tropical cyclone track forecasts.}
\textbf{a--f,} Forecast tracks (a,c,e) and track errors (b,d,f) for Haikui (a,b), Saola (c,d) and Bolaven (e,f). Forecasts for Haikui and Saola were initialized at 00 UTC on 29 August 2023 and evaluated over 168 and 144~h, respectively; the Bolaven forecast was initialized at 00 UTC on 9 October 2023 and evaluated over 168~h. Black lines show IBTrACS best tracks; orange, green and blue lines show Pangu-Weather-0.25\textdegree{}, NeuralGCM-0.7\textdegree{} and One2N-0.07\textdegree{}, respectively. Gold stars mark the initial positions, and track markers are spaced at 24-h intervals. Track error is the great-circle distance between the forecast cyclone centre and the best-track position at the same valid time. Case-study curves include occasional positions estimated by motion-based extrapolation when no centre was detected.
\textbf{g,} Median track errors for forecasts initialized in 2023. Shading indicates pointwise 95\% confidence intervals obtained from 2,000 bootstrap resamples, with tropical cyclones as the resampling clusters (Supplementary Section~\ref{subsec:tc_evaluation_protocol}). The effective sample size varies with lead time. At each lead time, verification includes only matched samples with a valid best-track position and a detected candidate in all three models. Forecast centres are obtained using the confidence-weighted tracking rule; purely extrapolated positions are excluded.
\textbf{h,} Distributions of paired track-error differences at 24, 48, 72, 96 and 120~h, using the same common-sample criterion as in g. Differences are defined as the comparator model's error minus the One2N-0.07\textdegree{} error; positive values indicate lower errors for One2N-0.07\textdegree{}. Orange and green distributions represent comparisons with Pangu-Weather-0.25\textdegree{} and NeuralGCM-0.7\textdegree{}, respectively. Shading shows kernel density estimates, and points show individual paired samples. Open squares and triangles indicate the medians of the Pangu-Weather-0.25\textdegree{} and NeuralGCM-0.7\textdegree{} comparisons, respectively.}
\label{fig:tc_tracking}
\end{figure*}

Across the three case studies, One2N-0.07\textdegree{} largely retained the track evolution predicted by NeuralGCM-0.7\textdegree{} while exhibiting smaller long-lead track errors than Pangu-Weather-0.25\textdegree{} (Fig.~\ref{fig:tc_tracking}a--f). For Haikui and Bolaven, the Pangu-Weather-0.25\textdegree{} tracks deviated markedly to the north and west of the best tracks, respectively, whereas One2N-0.07\textdegree{} more closely followed the observed cyclone motion. At 168~h, the One2N-0.07\textdegree{} track errors were 195.5 and 423.9~km, compared with 2,353.6 and 2,535.5~km for Pangu-Weather-0.25\textdegree{}, respectively. Saola showed a similar ranking, with One2N-0.07\textdegree{} close to NeuralGCM-0.7\textdegree{} and both outperforming Pangu-Weather-0.25\textdegree{} at long lead times. The close agreement between the One2N-0.07\textdegree{} and NeuralGCM-0.7\textdegree{} tracks is consistent with the retention of large-scale forecast patterns described above and shows similar track forecasts from the parent and refined models in these cases.

The 2023 sample supported the case-study findings (Fig.~\ref{fig:tc_tracking}g). Relative to Pangu-Weather-0.25\textdegree{}, One2N-0.07\textdegree{} reduced median track error by 16.5\%, 44.2\%, 58.4\% and 66.9\% at 24, 72, 120 and 168~h, respectively, with larger differences at longer lead times. At 168~h, median track error decreased from 1,024.8 to 339.4~km. Meanwhile, the median error curves for One2N-0.07\textdegree{} and NeuralGCM-0.7\textdegree{} remained close, extending the evidence for retention of the parent model's track performance beyond the three case studies.

Paired comparisons further distinguished these two relationships (Fig.~\ref{fig:tc_tracking}h). Across the five lead times shown, One2N-0.07\textdegree{} had lower track errors than Pangu-Weather-0.25\textdegree{} in 60.5\%--87.0\% of matched forecasts, indicating that the lower median errors were accompanied by improvements in a majority of individual cases. By contrast, the median paired differences between NeuralGCM-0.7\textdegree{} and One2N-0.07\textdegree{}, defined as the former's error minus the latter's, ranged from $-1.6$ to $0.7$~km. These paired comparisons show median track performance close to that of the parent model. On matched valid samples with a detected centre in all three models, One2N-0.07\textdegree{} also had lower median track  errors than the evaluated Pangu-Weather-0.25\textdegree{} configuration.

\section{Discussion}\label{sec4}
We introduce One2N, a recursive training framework that builds a high-resolution system from a single coarse baseline either an established predictor reused as it stands, or a coarse model trained at modest cost by adding successive refinement tiers. The framework retains the parent forecast operator and concentrates additional learning for finer grids in an independent residual module. In the two-level demonstration from NeuralGCM-0.7\textdegree{} to 0.07\textdegree{}, the combined system maintains broadly comparable skill to the parent model for representative tropospheric diagnostics and tropical cyclone tracks, while improving forecasts of key upper-atmospheric variables.

One2N separates reuse of the parent predictor from optimization of the added module. Unlike joint optimization of the parent and refinement modules, One2N freezes the parent parameters at each level and restricts gradient computation to the new module. The parent supplies a conditioning trajectory carrying information about large-scale evolution, and the residual module advances high-resolution residuals conditional on this trajectory. Gradients, optimizer states and intermediate activations required for backpropagation therefore need not be retained for the frozen parent. Training on a single GPU demonstrates the feasibility of this approach for the present global 0.07\textdegree{} configuration with modest computational resources. Refining the output grid therefore does not require re-optimizing the complete forecasting system, allowing prior training investment to support subsequent upgrades. The complete system still incurs the costs of parent forward integration, target-grid state storage and residual-network computation. How total cost scales with resolution and refinement depth therefore remains to be established through benchmarks with comparable data and forecast-skill requirements.

The vertical structure of the forecast gains provides insight into what the residual module learns. One2N-0.07\textdegree{} differs little from the parent for most tropospheric diagnostics, whereas improvements in upper-atmospheric geopotential and temperature coincide with substantial parent-model errors against both HRES-0.07\textdegree{} and ERA5-0.25\textdegree{}. Differences between the two analyses are smaller than the parent-model errors, and One2N-0.07\textdegree{} reduces those errors under either reference, indicating that the improvements extend beyond closer agreement with HRES-0.07\textdegree{} analyses as the training target. Improvements that emerge at short lead times and persist throughout the ten-day integration are also consistent with the correction of repeatable systematic errors. Incremental learning therefore combines spatial refinement with state correction, with benefits shaped jointly by the parent’s error structure and the supervision target. Tropical cyclone track results further indicate that the additional module retains the parent’s skill for this high-impact weather application while increasing output resolution. These vertical diagnostics support a contribution from systematic error correction; identifying the contribution of specific unresolved physical processes requires process-oriented evaluation.

The present evidence remains limited to one forecasting system, a single refinement level (\(N=2\), corresponding to a two-level model) and medium-range forecasting. This scope partly reflects the limited computational resources available for this study; deeper recursion and broader benchmarking remain future work. That baseline is a published model. Training a baseline from scratch and refining it is a design option that has not been benchmarked in this study, so the cost advantage of that route remains to be quantified. The recursive formulation allows further extension, but refinement depth and the resolution ratios between levels affect the learning task, error propagation and cumulative cost and therefore require joint evaluation. The choice of training targets also constrains interpretation.  HRES-0.07\textdegree{} is an analysis product that combines observations with model background information \cite{bonavita2016evolution}, so the learned residual may contain contributions from scale differences, parent-model biases and analysis-system characteristics. Verification against independent observations would help establish the practical forecast value of these corrections. The generality of the state interface does not imply direct transfer of a particular residual module across grids. The Swin-Unet architecture and spatial rearrangement used here are designed for a structured latitude–longitude grid; refinement on unstructured grids such as those of MPAS and ICON would require state mappings and residual parameterizations compatible with the mesh topology \cite{skamarock2012multiscale,zangl2015icon}.

At the system level, One2N treats a running forecast operator as a reusable component. Foundation models such as Aurora demonstrate the value of shared pretrained knowledge for task adaptation \cite{bodnar2025foundation}, while generative downscaling offers ways to combine coarse- and fine-scale information \cite{hess2025fast,mardani2025residual}. One2N offers a complementary strategy: with compatible state and temporal interfaces, established physics-based, data-driven or hybrid predictors can, in principle, be combined with independently designed refinement modules. In physical modelling, increasing resolution changes the partition between resolved motions and parameterized processes \cite{schar2020kilometer}. By treating scale-dependent state correction as a separate learning task, One2N provides a way to investigate adaptation across resolutions while retaining the underlying predictor. This allows accumulated expertise in dynamical modelling, training, calibration and verification to continue contributing to new systems, while enabling independent assessment of improvements and degradations introduced by the additional modules.

Building on this framework, further extensions could address resource allocation and cross-scale feedback. Local refinement could concentrate additional computation in regions where detailed forecasts are needed. With architectures that support regional inputs, sufficient spatial context and exchange of boundary information, domain decomposition could reduce per-device activation memory requirements. If such regional modules were combined to cover the globe, whether they could form a consistent global system would still require evaluation of propagation across boundaries and autoregressive stability. A second direction is to feed fine-scale corrections back to the parent model, allowing investigation of interactions between scales that are absent from the current one-way conditioning. Hybrid modelling studies show that such coupling affects accuracy and integration stability \cite{brenowitz2020interpreting,kochkov2024neural}; feedback can alter the actual parent trajectory even when its parameters remain fixed. Parameter freezing alone does not guarantee conservation properties, balance relationships among variables or long-term statistics of the combined fields. These properties require independent evaluation in full forecast integrations.

The One2N demonstration shows that the additional learning required for resolution refinement can be separated from retraining an established predictor, allowing existing predictive capability to serve as a computational foundation for subsequent system development. By freezing the parent model and independently learning scale-dependent corrections, the framework provides a modular route to global high-resolution forecasting that builds on accumulated scientific and engineering expertise, and offers a training strategy for multiscale Earth system modelling under resource constraints. Because the baseline is trained at coarse resolution, the entry cost scales with the coarse grid rather than with the target grid; the same strategy therefore serves groups that already operate a forecasting system and groups that do not.

\section{Methods}\label{sec2}

\subsection{The One2N training framework}

The One2N training framework decomposes the construction of a target-resolution model into a frozen baseline (tier 0) plus \(N\) successive refinement tiers  (tiers \(0,\ldots,N-1\)) (Fig.~\ref{fig:one2n_framework}a). At tier \(i=0,\ldots,N-1 (N \geq 1)\), the predictor \(\mathrm{One2N}^{(i)}\) is formed by augmenting the preceding coarser-resolution predictor \(\mathrm{One2N}^{(i-1)}\) with a tier-specific incremental operator \(\Delta\mathrm{One2N}^{(i)}\):

\begin{equation}\label{eq1}
\mathrm{One2N}^{(i)}=\mathrm{One2N}^{(i-1)}+\Delta\mathrm{One2N}^{(i)},
\end{equation}
where \(\mathrm{One2N}^{(0)}\) denotes the baseline coarse-resolution model. At each tier, the complete parent forecast operator is frozen, and only the incremental operator is optimized. The addition in Eq.~\eqref{eq1} is implemented at the state level by combining the parent forecast mapped to the target grid with the predicted residual. The framework follows a four-step procedure inspired by multigrid methods\cite{zhang2022integral}, organizing error correction through successive resolution refinements. This study evaluates a single refinement tier (\(N=2\)), corresponding to a two-level model comprising the frozen parent and one refinement tier.

\textbf{Step 1: Baseline model construction.}
The baseline model, \(\mathrm{One2N}^{(0)}\), can be constructed by training a neural network on the coarsest-resolution dataset, typically derived from atmospheric reanalysis; because that grid is coarse, the training is inexpensive. Candidate architectures include residual networks\cite{he2016deep}, U-Net variants\cite{ronneberger2015unet}, graph neural networks\cite{battaglia2018relational} and attention-based models\cite{vaswani2017attention}. Alternatively, an existing coarse-resolution forecasting system can be adopted directly as \(\mathrm{One2N}^{(0)}\), including numerical, data-driven and hybrid predictors\cite{kashinath2021physics}. In either case, the parent must provide compatible state variables at the times required by the incremental operator. The two routes differ in cost: reusing an existing system adds only the tier's own training, whereas a new baseline is trained on a much coarser grid typically hundreds of kilometres versus a target resolution that can reach kilometre scales and thus with orders of magnitude fewer grid points.

In this study, \(\mathrm{One2N}^{(0)}\) is instantiated as NeuralGCM-0.7\textdegree{},  a hybrid model that combines a differentiable atmospheric dynamical core with learned physical parameterizations\cite{kochkov2024neural}.

\textbf{Step 2: Temporal integration of the parent coarse-resolution model.}
For each training sample at tier \(i\), the initial high-resolution analysis \(x^*_{h,t_0}\) is mapped to the parent grid using a fixed operator \(\mathcal D_i\), giving \(x_{l,t_0}=\mathcal D_i(x^*_{h,t_0})\). Here, \(h\) and \(l\) denote the target and parent grids, respectively, and an asterisk denotes a reference analysis. Starting from this initial state, the frozen parent model is integrated forward to generate the trajectory

\begin{equation}\label{eq:parent_trajectory}
x_{l,t_j}=\mathrm{One2N}^{(i-1)}_{\delta t}
\left(x_{l,t_{j-1}}\right),
\qquad j=1,\ldots,N_\tau,
\end{equation}
at times \(t_j=t_0+j\delta t\), where \(j=0,\ldots,N_\tau\) and \(N_\tau=\tau/\delta t\). Here, \(\delta t\) is the interval between successive residual updates, and \(\tau\) is the interval between supervised high-resolution states. The subscript \(\delta t\) denotes advancement of the parent model over this interval, which may comprise multiple internal integration steps. In the present implementation, the parent is integrated continuously in its native state representation, with decoded pressure-level states supplied to the correction module every \(\delta t=2\)~h. The parent-step notation represents advancement of the continuously maintained native model state; decoded pressure-level fields are not re-encoded at each correction step. High-resolution reference analyses provide supervision at \(\tau=6\)~h, giving \(N_\tau=3\).
No intermediate analysis updates are applied to the parent trajectory. Initialization inputs and the surface-forcing configuration are described in Supplementary Sections~\ref{subsec:supp_neuralgcm} and~\ref{subsec:forecast_initialization}.

\textbf{Step 3: Formulation of the incremental model.}
The incremental model predicts a correction to the parent forecast using information from the preceding high-resolution state and the parent states before and after each forecast interval. We parameterize this dependence through the difference between the preceding high-resolution state and the mapped parent state, together with the parent temporal increment:

\begin{equation}\label{eq2}
\widehat{\Delta x}_{h,t_j}=
\Delta\mathrm{One2N}^{(i)}_{\theta_i}
\left(\widehat{x}_{h,t_{j-1}}-\mathcal I_i\left(x_{l,t_{j-1}}\right),\;
x_{l,t_j}-x_{l,t_{j-1}}\right).
\end{equation}
Here, \(\widehat{x}_{h,t_{j-1}}\) is the high-resolution forecast at the preceding time, \(x_{l,t_{j-1}}\) is the parent state at that time, and \(x_{l,t_j}\) is the parent forecast obtained in Step 2. The fixed operator \(\mathcal I_i\) maps a parent state to the target grid, and \(\theta_i\) denotes the trainable parameters of the incremental model.

For compact notation, we define the parent temporal increment as

\begin{equation}\label{eq3}
\Delta x_{l,t_j}=x_{l,t_j}-x_{l,t_{j-1}},
\end{equation}
and the preceding high-resolution residual as

\begin{equation}\label{eq4}
\widehat{\Delta x}_{h,t_{j-1}}=
\widehat{x}_{h,t_{j-1}}-\mathcal I_i\left(x_{l,t_{j-1}}\right).
\end{equation}
With these definitions, Eq.~\eqref{eq2} becomes

\begin{equation}\label{eq5}
\widehat{\Delta x}_{h,t_j}=
\Delta\mathrm{One2N}^{(i)}_{\theta_i}\left(
\widehat{\Delta x}_{h,t_{j-1}},\Delta x_{l,t_j}\right),
\qquad j=1,\ldots,N_\tau.
\end{equation}

The recurrence is initialized with the reference state \(\widehat{x}_{h,t_0}=x^*_{h,t_0}\), giving

\begin{equation}\label{eq:initial_residual}
\widehat{\Delta x}_{h,t_0}:=
\Delta x^*_{h,t_0}=
x^*_{h,t_0}-\mathcal I_i\left(x_{l,t_0}\right).
\end{equation}
The operator is applied autoregressively, using each predicted residual as the input to the next update (Fig.~\ref{fig:one2n_framework}b).

At each forecast time, the predicted residual is added to the parent forecast mapped to the target grid to reconstruct the high-resolution state:

\begin{equation}\label{eq:state_reconstruction}
\widehat{x}_{h,t_j}=
\mathcal I_i\left(x_{l,t_j}\right)+\widehat{\Delta x}_{h,t_j}.
\end{equation}
The parent trajectory evolves independently of these residual corrections, providing one-way conditioning.

\textbf{Step 4: Training of the incremental model.}
The incremental operator is parameterized by a 3D Swin-Unet and trained using the high-resolution dataset and the parent trajectories generated in Step 2. For each training sample \(k\), the recurrence in Eq.~\eqref{eq5} is unrolled for \(N_\tau\) steps with shared parameters \(\theta_i\), starting from the initial residual defined in Eq.~\eqref{eq:initial_residual}.

Training minimizes the discrepancy between the terminal predicted residual produced by these nested applications and the corresponding reference residual:

\begin{equation}\label{eq6}
\begin{aligned}
\mathcal L^{(i)}(\theta_i)={}&
\frac{1}{K}\sum_{k=1}^{K}
\Bigl\|\Delta\mathrm{One2N}^{(i)}_{\theta_i}
\Bigl(\cdots\Delta\mathrm{One2N}^{(i)}_{\theta_i}
\bigl(\Delta x^*_{h,t_0,k},\Delta x_{l,t_1,k}
\bigr)\cdots,\\
&\qquad
\Delta x_{l,t_{N_\tau},k}
\Bigr)-\Delta x^*_{h,t_{N_\tau},k}
\Bigr\|_{w}^{2},
\end{aligned}
\end{equation}
where \(K\) is the number of training samples. The nested expression contains \(N_\tau\) applications of the same operator, with the parent increment \(\Delta x_{l,t_j,k}\) supplied at step \(j\). The terminal reference residual is

\begin{equation}\label{eq:terminal_residual_target}
\Delta x^*_{h,t_{N_\tau},k}=
x^*_{h,t_{N_\tau},k}-\mathcal I_i\left(x_{l,t_{N_\tau},k}\right).
\end{equation}
Here, \(\|\cdot\|_w^2\) denotes MSE in the standardized training representation after restoration to the target-grid arrangement and multiplication of both prediction and target by \(\cos\phi\). The effective weight on squared errors is therefore \(\cos^2\phi\) (Supplementary Eq.~\eqref{eq:supp_terminal_training_loss}).

The loss is evaluated only at \(t_{N_\tau}\) and differentiated through all \(N_\tau\) residual updates. The parent trajectory remains detached from the optimization graph, and only \(\theta_i\) is updated. After training, the learned incremental operator is combined with the frozen parent according to Eq.~\eqref{eq1}. The resulting predictor can serve as the parent for a subsequent refinement tier, for which Steps 2--4 are repeated.

\subsection{Network architecture and training}
The incremental module is trained on six-hourly ECMWF operational HRES-0.07\textdegree{} pressure-level analyses from the NCAR archive (dataset d113001, formerly ds113.1)\cite{ecmwf2016hresdata}. The target variables are zonal wind, meridional wind, temperature, specific humidity and geopotential on 25 pressure levels. Data from 2016--2020 are used for training and 2021 for validation, while held-out evaluation uses cases from 2023. Standardization statistics are estimated from the training data and kept fixed during validation and testing. Further dataset specifications and descriptions of the verification datasets are provided in Supplementary Section~\ref{sec:supp_note1}.

The incremental operator \(\Delta\mathrm{One2N}^{(i)}_{\theta_i}\) is parameterized by a three-dimensional Swin-Unet adapted for cross-resolution residual evolution\cite{liu2021swin,cao2023swinunet} (Fig.~\ref{fig:one2n_framework}c). The architecture combines value-preserving spatial-channel folding, nonlinear fusion of parent-model increments and fine-grid residual states, and hierarchical reconstruction to implement the conditional update in Eq.~\eqref{eq5}.

\textbf{Spatial--channel folding of high-resolution residuals.}
A fixed, invertible rearrangement \(\mathcal P_i\) reorganizes horizontal tiles of the high-resolution residual into channel groups while retaining pressure levels as a separate axis. In the present implementation, each \(2560\times5120\) field is divided into 100 non-overlapping \(256\times512\) tiles. For five variables on 25 pressure levels, the rearranged residual has shape \((500,25,256,512)\) , where \(500 = 5\) variables \(\times 100\) tiles. The folding operation preserves all input field values before learned embedding. Fine-grid information is therefore carried by additional channels, while the horizontal array used for subsequent patch embedding remains \(256\times512\). The five parent-increment channels and the folded residual form the joint input

\begin{equation}\label{eq:increment_network_input}
\mathbf X_{i,j}
=
\operatorname{Concat}_{c}
\left[
\widetilde{\Delta x}_{l,t_j},
\mathcal P_i\left(\widehat{\widetilde{\Delta x}}_{h,t_{j-1}}\right)
\right].
\end{equation}
Here, \(\operatorname{Concat}_{c}\) denotes channel-wise concatenation, and \(\mathbf X_{i,j}\) has shape \((B,505,25,256,512)\), where \(B\) is the batch size. Tildes denote standardization using the fixed statistics in Supplementary Eq.~\eqref{eq:supp_standardization}. The first five channels contain the parent increments, followed by the 500 rearranged residual channels. The fixed tile-to-channel assignment allows the network to learn the relationship between parent increments and rearranged fine-grid residuals.

\textbf{Fusion of temporal increments and spatial residuals.}
The parent increment describes the resolved state change over the forecast interval, while the high-resolution residual describes the departure from the mapped parent state. Joint three-dimensional patch embedding and subsequent Transformer blocks allow the residual update to depend nonlinearly on both representations. This couples the correction of the current fine-grid state to the evolving parent forecast. Temporal information enters through successive parent increments and autoregressive residual updates; the attention axes correspond to pressure-level groups and the two horizontal token coordinates.

\textbf{Hierarchical attention and multiscale residual reconstruction.}
The backbone alternates window-based multi-head self-attention (W-MSA) and shifted-window multi-head self-attention (SW-MSA), combining interactions within windows with information exchange across neighbouring windows\cite{liu2021swin}. Patch merging aggregates context at progressively coarser horizontal resolutions, and patch expansion restores feature resolution in the decoder. One skip connection concatenates encoder features with decoder features at the matching resolution, followed by linear fusion and further Swin processing. This path provides reconstruction with features from less downsampled stages alongside the contextual information accumulated in deeper layers. Layer normalization, multilayer perceptrons and residual connections are used within the Swin blocks.

A three-dimensional patch-recovery layer produces the rearranged output \(\widehat{\mathbf Y}_{i,j}\). The channel-selection operator \(\Pi_i\) extracts the predicted residual channels, and the inverse rearrangement restores the target-grid residual:

\begin{equation}\label{eq:increment_network_output}
\widehat{\Delta x}_{h,t_j}=
\mu_h+\sqrt{\sigma_h^2+\epsilon}\,\odot\,
\mathcal P_i^{-1}\left[
\Pi_i\left(\widehat{\mathbf Y}_{i,j}\right)
\right].
\end{equation}
Here, \(\mu_h\), \(\sigma_h^2\) and \(\epsilon\) are the residual standardization statistics defined in Supplementary Eq.~\eqref{eq:supp_standardization}, and \(\odot\) denotes elementwise multiplication. The residual is restored to physical units before combination with the mapped parent forecast according to Eq.~\eqref{eq:state_reconstruction}.

\textbf{Three-dimensional positional encoding.}
The learned relative position bias depends on token offsets along the vertical and two horizontal axes within each attention window. Different directions and separations can therefore receive distinct attention biases, providing a representation of direction-dependent spatial relationships in token space. Absolute positional embeddings additionally identify positions on the token grid. The complete network configuration is provided in Supplementary Table~\ref{tab:supp_network_configuration}.

\textbf{Network configuration and optimization.}
Optimization of the multi-step objective in Eq.~\eqref{eq6} uses SGD with an initial learning rate of \(10^{-3}\), subsequently reduced to \(5\times10^{-4}\) and \(10^{-4}\) in response to plateaus in the training loss (Supplementary Section~\ref{subsec:supp_optimization_rollout}), and a batch size of one. Parameters are updated once per 6~h training window, after backpropagation through all three 2~h residual updates. Training the incremental module takes approximately four weeks on a single NVIDIA A800 GPU with 80 GB of memory. Further optimization details are provided in Supplementary Section~\ref{subsec:supp_optimization_rollout}.

\section*{Data availability}
\begingroup\raggedright
ERA5-0.25\textdegree{} pressure-level reanalysis data are available from the Copernicus Climate Data Store (\url{https://doi.org/10.24381/cds.bd0915c6})\cite{c3s2018era5data}. The HRES-0.07\textdegree{} analyses used here are distributed within the ECMWF IFS High-Resolution Operational Forecasts collection at the NCAR Geoscience Data Exchange, dataset d113001 (formerly ds113.1; \url{https://doi.org/10.5065/D68050ZV}). IBTrACS v04r01 is available from NOAA's National Centers for Environmental Information (\url{https://www.ncei.noaa.gov/data/international-best-track-archive-for-climate-stewardship-ibtracs/v04r01/}). Access to the source datasets is subject to the respective providers' terms. 
\endgroup

\section*{Code availability}
NeuralGCM-0.7\textdegree{} code and checkpoint documentation are available at \url{https://github.com/neuralgcm/neuralgcm}; the Pangu-Weather-0.25\textdegree{} implementation is available at \url{https://github.com/198808xc/Pangu-Weather}.

\begin{backmatter}
\section*{Acknowledgements}
This work was partially supported by the National Natural Science Foundation of China (Grant No.  42635011) and the National Key Research and Development Project of China (Grant No. 2024YFF0809104). 
We thank the National Large Scientific and Technological Infrastructure ``Earth System Numerical Simulation Facility'' (\url{https://cstr.cn/31134.02.EL}) for technical support.

\end{backmatter}

\clearpage
\bibliographystyle{sn-nature}
\bibliography{one2n-references}

\raggedbottom
\clearpage
\setcounter{page}{1}
\renewcommand{\thepage}{S\arabic{page}}
\setcounter{section}{0}
\setcounter{subsection}{0}
\setcounter{subsubsection}{0}

\setcounter{secnumdepth}{2}

\renewcommand{\thesection}{\arabic{section}}
\renewcommand{\thesubsection}{\thesection.\arabic{subsection}}
\renewcommand{\thesubsubsection}{\thesubsection.\arabic{subsubsection}}

\setcounter{figure}{0}
\setcounter{table}{0}
\setcounter{equation}{0}
\renewcommand{\theequation}{S\arabic{equation}}
\renewcommand{\theHequation}{supp.\arabic{equation}}
\renewcommand{\figurename}{Supplementary Fig.}
\renewcommand{\tablename}{Supplementary Table}

\renewcommand{\thefigure}{S\arabic{figure}}
\renewcommand{\thetable}{S\arabic{table}}

\renewcommand{\theHsection}{supp.\arabic{section}}
\renewcommand{\theHsubsection}{\theHsection.\arabic{subsection}}
\renewcommand{\theHsubsubsection}{\theHsubsection.\arabic{subsubsection}}

\renewcommand{\theHfigure}{supp.\arabic{figure}}
\renewcommand{\theHtable}{supp.\arabic{table}}

\startcontents[supp]

\begin{center}
    {\Large\bfseries One2N: Recursive refinement breaks the resolution barrier for Earth-system AI\par}
    \vspace{8mm}
    {\Huge\bfseries Supplementary Information\par}
\end{center}

\vspace{1cm}

\section*{Supplementary Contents}
\makeatletter
\let\OneTwoNSavedBibContext\the@ipfilectr
\makeatother
\printcontents[supp]{}{1}[2]{}
\makeatletter
\global\let\the@ipfilectr\OneTwoNSavedBibContext
\makeatother

\clearpage

\section{Datasets and preprocessing}
\label{sec:supp_note1}

Three datasets supported the experiments. HRES-0.07\textdegree{} analyses provided the high-resolution training
targets and primary verification reference, ERA5-0.25\textdegree{} provided an alternative analysis reference, and IBTrACS best tracks were used for tropical-cyclone verification. The datasets and their preparation are described below.

\subsection{Datasets}
\label{sec:supp_datasets}

\subsubsection{HRES-0.07\textdegree{} analyses}
\label{subsec:hres_dataset}

ECMWF operational high-resolution pressure-level analyses were obtained from the NSF NCAR archive d113001 (formerly ds113.1)\cite{ecmwf2016hresdata}. HRES-0.07\textdegree{} denotes the archived regular Gaussian grid with 2560 latitude nodes and 5120 uniformly spaced longitudes, referred to here by its nominal grid spacing of 0.07\textdegree{}. Five variables on 25 pressure levels define the One2N-0.07\textdegree{} target state (Table~\ref{tab:variables}). Analyses at 0000, 0600, 1200 and 1800 UTC were used, with 2016--2020 allocated to training, 2021 to validation and checkpoint selection, and 2023 to forecast evaluation. HRES-0.07\textdegree{} supplied the correction-module targets and the primary verification reference.

\begin{table}[htbp]
\centering
\small
\setlength{\tabcolsep}{5pt}
\renewcommand{\arraystretch}{1.12}
\caption{Variables defining the One2N-0.07\textdegree{} target state. Each variable is retained on 25 pressure levels: 1, 2, 3, 5, 7, 10, 20, 30, 50, 70, 100, 150, 200, 250, 300, 400, 500, 600, 700, 800, 850, 900, 925, 950 and 1000~hPa. ERA5-0.25\textdegree{}-based diagnostics use the same variables and levels.}
\label{tab:variables}
\begin{tabular*}{\textwidth}{@{\extracolsep{\fill}}lccccc@{}}
\hline
Type & Variable name & Short name & ECMWF Parameter ID & Unit & Role\\
\hline
Atmospheric & Geopotential & \(Z\) & 129 & \(\mathrm{m^2\,s^{-2}}\) & Input / Predicted \\
Atmospheric & Temperature & \(T\) & 130 & \(\mathrm{K}\) & Input / Predicted \\
Atmospheric & Zonal wind & \(U\) & 131 & \(\mathrm{m\,s^{-1}}\) & Input / Predicted \\
Atmospheric & Meridional wind & \(V\) & 132 & \(\mathrm{m\,s^{-1}}\) & Input / Predicted \\
Atmospheric & Specific humidity & \(Q\) & 133 & \(\mathrm{kg\,kg^{-1}}\) & Input / Predicted \\
\hline
\end{tabular*}

\end{table}

\subsubsection{ERA5-0.25\textdegree{} reanalysis}
\label{subsec:era5_dataset}

ERA5-0.25\textdegree{} pressure-level reanalysis fields\cite{hersbach2020era5} were obtained from the Copernicus Climate Data Store on a regular 0.25\textdegree{} latitude-longitude grid\cite{c3s2018era5data}. The variables and levels in Table~\ref{tab:variables} were selected from the 37 available pressure levels, and the hourly fields were sampled at the same four synoptic times as HRES-0.07\textdegree{}. ERA5-0.25\textdegree{} provided an alternative verification reference and contemporaneous analysis fields for the HRES-0.07\textdegree{}--ERA5-0.25\textdegree{} diagnostics. Because the parent model was trained on ERA5-0.25\textdegree{}-derived states, this reference also allowed its forecast errors to be assessed against the reanalysis associated with its training data. ERA5-0.25\textdegree{} supplied auxiliary cloud fields for NeuralGCM-0.7\textdegree{} initialization, but was not a supervisory target for the One2N-0.07\textdegree{} correction module.

\subsubsection{IBTrACS best tracks}
\label{subsec:ibtracs_dataset}

Tropical-cyclone centre positions were obtained from version v04r01 of the International Best Track Archive for Climate Stewardship (IBTrACS), distributed by NOAA's National Centers for Environmental Information\cite{knapp2010ibtracs,gahtan2024ibtracsdata}. Best-track positions at 0000, 0600, 1200 and 1800 UTC were paired with forecast centres at the same valid times.

\subsection{Preprocessing}
\label{sec:supp_preprocessing}

\subsubsection{Horizontal remapping}
\label{subsec:conservative_remap}

\begingroup\raggedright
Horizontal grid transfers used first-order conservative remapping\cite{jones1999conservative}, implemented with Dinosaur's \texttt{horizontal\_interpolation.\allowbreak ConservativeRegridder}. For a target cell \(j\) fully covered by valid source cells,\par

\endgroup
\begin{equation}
y_j=\sum_i \alpha_{ij}x_i,
\qquad
\alpha_{ij}=\frac{A_{ij}}{A_j},
\qquad
\sum_i\alpha_{ij}=1,
\label{eq:supp_conservative_remap}
\end{equation}
where \(x_i\) and \(y_j\) are source- and target-cell values, \(A_{ij}\) is the overlap area of source cell \(i\) and target cell \(j\), and \(A_j\) is the target-cell area. Remapping was applied separately to each variable, pressure level and valid time. Fields already on the required grid were retained without additional remapping. Scores were calculated from the remapped fields at each valid time, before aggregation across forecasts.

\subsubsection{Verification fields}
\label{subsec:verification_grids}

Forecasts and reference analyses were matched by variable, pressure level and valid time. Verification used the pressure levels shared by the participating systems, without vertical interpolation. The primary evaluation compared all forecasts with HRES-0.07\textdegree{} analyses on the HRES-0.07\textdegree{} grid. The alternative evaluation compared all forecasts with ERA5-0.25\textdegree{} on the regular 0.25\textdegree{} grid.

Cross-reference diagnostics also used the common 0.25\textdegree{} grid. HRES-0.07\textdegree{} analyses and the One2N-0.07\textdegree{} and NeuralGCM-0.7\textdegree{} forecasts were remapped to this grid so that forecast errors against either analysis and HRES-0.07\textdegree{}--ERA5-0.25\textdegree{} differences were evaluated at the same horizontal resolution. For the cross-reference diagnostics, all comparisons used the intersection of valid cells across HRES-0.07\textdegree{}, ERA5-0.25\textdegree{}, NeuralGCM-0.7\textdegree{} and One2N-0.07\textdegree{} on the common 0.25\textdegree{} grid. The ERA5-0.25\textdegree{}-based evaluation changed both the reference analysis and the verification grid; the fixed-grid cross-reference diagnostics compared errors against either analysis under the same spatial representation.

\section{Model configurations and experimental protocol}
\label{sec:supp_note2}

NeuralGCM-0.7\textdegree{} was used as the parent model and coarse-resolution baseline, and Pangu-Weather-0.25\textdegree{} as an external AI benchmark. The single-level One2N-0.07\textdegree{} configuration was evaluated against both models to assess the effects of refinement and its performance relative to an established data-driven forecasting system.

\subsection{NeuralGCM-0.7\textdegree{}}
\label{subsec:supp_neuralgcm}

NeuralGCM-0.7\textdegree{} couples an atmospheric dynamical core to learned physical parameterizations\cite{kochkov2024neural}. Its encoder maps pressure-level inputs to an internal sigma-coordinate state, which is advanced by the dynamical core and learned parameterizations; the decoder returns pressure-level fields. The pretrained pressure-level interface includes horizontal winds, temperature, specific humidity, geopotential, specific cloud liquid water content and specific cloud ice water content on 37 levels. Sea surface temperature and sea-ice cover provide surface forcing.

The pretrained 0.7\textdegree{} checkpoint was used without parameter updates. Atmospheric initial fields were obtained from HRES-0.07\textdegree{}, except for cloud liquid water and cloud ice, which were supplied from ERA5-0.25\textdegree{} at the same initialization time because they were unavailable in the HRES-0.07\textdegree{} analyses used here. The five HRES-0.07\textdegree{} atmospheric variables were interpolated from 25 to 37 pressure levels using linear interpolation in logarithmic pressure for NeuralGCM-0.7\textdegree{} initialization. Sea surface temperature and sea-ice cover were obtained from HRES-0.07\textdegree{} and updated at 7-day intervals. Surface-forcing updates used HRES-0.07\textdegree{} forecasts initialized at the forecast start time.

After encoding at initialization, the parent was integrated continuously in its native state representation, with decoded pressure-level states supplied to the correction module every 2~h. This interval defines the coupling interface, rather than the internal numerical time step. The standalone NeuralGCM-0.7\textdegree{} baseline used the same checkpoint, initial inputs and surface-forcing configuration as the parent component of  One2N-0.07\textdegree{}. Verification selected the five target variables on the 25 pressure levels in Table~\ref{tab:variables}, without vertical interpolation.

\subsection{Pangu-Weather-0.25\textdegree{}}
\label{subsec:supp_pangu}

Pangu-Weather-0.25\textdegree{} uses a three-dimensional Earth-specific Transformer to predict upper-air and surface fields on a regular 0.25\textdegree{} grid\cite{bi2023accurate}. Its upper-air state comprises the five variables in Table~\ref{tab:variables} on 13 pressure levels: 50, 100, 150, 200, 250, 300, 400, 500, 600, 700, 850, 925 and 1000~hPa. Its surface state comprises mean sea-level pressure, 10~m zonal and meridional winds, and 2~m temperature.

The pretrained 3~h checkpoint was applied autoregressively, with predicted upper-air and surface states supplied to the next application. All initial fields were obtained from HRES-0.07\textdegree{} and mapped to the required input grid. All Pangu-Weather-0.25\textdegree{} results reported here refer to this repeated 3~h configuration, and gridded comparisons used the 13 pressure levels shared with One2N-0.07\textdegree{}.

\subsection{One2N-0.07\textdegree{}}
\label{subsec:supp_one2n_configuration}

One2N-0.07\textdegree{} combined the frozen NeuralGCM-0.7\textdegree{} parent with a separately trained 3D Swin-Unet correction module. Only one refinement level was evaluated (\(N=2\)). The correction module advanced the target-grid residual every 2~h, conditioned on the parent temporal increment and the preceding residual. Adding the predicted residual to the mapped parent forecast reconstructed the five-variable state on 25 pressure levels. The parent trajectory evolved independently of the residual corrections.

The correction module was trained using HRES-0.07\textdegree{} analysis targets over 6~h windows, corresponding to three residual updates. The loss was evaluated at the final update and differentiated through all three applications of the correction module. During forecast evaluation, predicted residuals were propagated autoregressively beyond the training window without replacement by reference residuals. Network and optimization details are given in Supplementary Section~\ref{sec:supp_note3}.

\subsection{Initialization and evaluation}
\label{subsec:forecast_initialization}

For all three systems, the atmospheric variables shared by the models were initialized from the HRES-0.07\textdegree{} analysis at the same start time, after the transformations required by each input representation. One2N-0.07\textdegree{} used the target-grid analysis to initialize its high-resolution state. The NeuralGCM-0.7\textdegree{} baseline and the One2N-0.07\textdegree{} parent received identical auxiliary cloud inputs and surface forcings. Atmospheric states were not reinitialized after the forecast start time.

Global gridded evaluation used 120 forecasts initialized at randomly selected analysis times in 2023. Each forecast extended to 240~h, with the same start times used for all three models. Verification sampled the common lead times 12, 24, \ldots, 240~h directly from the forecast sequences, without temporal interpolation. Thus, the 2~h correction interval and the 3~h Pangu-Weather-0.25\textdegree{} prediction interval did not alter the valid times included in the comparison. Comparisons with NeuralGCM-0.7\textdegree{} covered five variables on 25 levels; comparisons with Pangu-Weather-0.25\textdegree{} covered five variables on 13 common levels.

Tropical-cyclone evaluation used a separate sample of 185 cyclone initialization pairs for 84 cyclones in 2023, with matched samples determined separately at each lead time. The case studies were initialized at 0000 UTC on 29 August 2023 for Haikui and Saola, and at 0000 UTC on 9 October 2023 for Bolaven. Verification extended to 168~h for Haikui and Bolaven and 144~h for Saola. The tracking procedure and sample-selection criteria are described in Supplementary Section~\ref{sec:supp_note7}.

\section{Implementation details of One2N-0.07\textdegree{}}
\label{sec:supp_note3}

\subsection{Residual representation and correction network}
\label{subsec:supp_correction_network}

Let \(x_{l,t}\) denote the five decoded parent variables on the 25 target pressure levels, and let \(x^*_{h,t}\) denote the corresponding HRES-0.07\textdegree{} analysis. The mapping \(\mathcal I\) transfers the parent state to the target horizontal grid. At the correction times \(t_j=t_0+j\delta t\), with \(\delta t=2\)~h, the parent temporal increment is

\begin{equation}
\Delta x_{l,t_j}=x_{l,t_j}-x_{l,t_{j-1}}.
\label{eq:supp_parent_increment}
\end{equation}
At a supervised valid time \(t\), the reference residual is

\begin{equation}
\Delta x^*_{h,t}=x^*_{h,t}-\mathcal I(x_{l,t}).
\label{eq:supp_residual_target}
\end{equation}
Thus, the correction target contains the difference between the high-resolution analysis and the mapped parent forecast, including both spatial structure and state error.

For network input, each \(2560\times5120\) residual field was partitioned into a \(10\times10\) array of contiguous, non-overlapping \(256\times512\) tiles. A fixed rearrangement \(\mathcal P\) stacked the 100 tiles along the channel axis within each variable, preserving the 25-level axis. Tiles were ordered by the first horizontal tile index and then the second. The rearranged residual had shape \((B,500,25,256,512)\) and was concatenated after the five parent-increment channels, giving an input of shape \((B,505,25,256,512)\). Here \(B\) is the batch size. Standardization was applied before rearrangement, as described below.

The correction network was a three-dimensional adaptation of Swin-Unet\cite{cao2023swinunet,liu2021swin}, implemented in PyTorch. Patch embedding produced a \(5\times16\times32\) token grid with 128 features per token. Two encoder stages and two decoder stages used windowed and shifted-window attention, with horizontal downsampling and upsampling between resolutions. A skip connection concatenated the patch-embedding features with the upsampled decoder features before the final decoder stage; a linear projection fused these features. The configuration is given in Table~\ref{tab:supp_network_configuration}.

Patch recovery returned 505 channels. The final 500 channels were selected and restored by \(\mathcal P^{-1}\) to form the next standardized residual on the target grid. After conversion to physical units, the predicted residual was added to the mapped parent forecast to reconstruct the high-resolution state. The same network was reused at every correction step.

\begin{table}[htbp]
\centering
\small
\setlength{\tabcolsep}{5pt}
\renewcommand{\arraystretch}{1.12}
\caption{Configuration of the 3D Swin-Unet correction module. Patch, window, downsampling and upsampling dimensions follow the pressure-level axis and the two horizontal axes of the rearranged representation. Encoder and decoder depths specify the numbers of attention blocks in their respective stages.}
\label{tab:supp_network_configuration}
\begin{tabular*}{\textwidth}{@{\extracolsep{\fill}}ll@{}}
\hline
Parameter & Configuration \\
\hline
Patch size & \((5,16,16)\) \\
Initial embedding dimension & 128 \\
Encoder stage depths & \((2,4)\) \\
Decoder stage depths & \((4,2)\) \\
Attention heads & 4 per stage \\
Attention window & \((5,4,4)\) \\
Downsampling and upsampling & \((1,2,2)\) \\
MLP expansion ratio & 4 \\
Normalization & Layer normalization; patch normalization enabled \\
Position encoding & Learned absolute embeddings and relative position biases \\
Maximum drop-path rate & 0.1 \\
\hline
\end{tabular*}

\end{table}

\subsection{Standardization and optimization}
\label{subsec:supp_optimization_rollout}

Parent temporal increments and target-grid residuals used separate standardization statistics. For each variable and pressure level, means and variances were estimated from valid 2016--2020 training data using cosine-latitude spatial weighting. Parent-increment statistics pooled the three consecutive 2~h increments within the 6~h training windows; residual statistics used the reference residuals at initialization and at the supervised 6~h endpoints. Standardization was applied as

\begin{equation}
\begin{aligned}
\widetilde{\Delta x}_l
&=
\frac{\Delta x_l-\mu_l}{\sqrt{\sigma_l^2+\epsilon}},\\
\widetilde{\Delta x}_h
&=
\frac{\Delta x_h-\mu_h}{\sqrt{\sigma_h^2+\epsilon}},
\end{aligned}
\label{eq:supp_standardization}
\end{equation}
where \((\mu_l,\sigma_l^2)\) and \((\mu_h,\sigma_h^2)\) refer to the parent increments and high-resolution residuals, respectively. The stabilizing constant \(\epsilon\) was set to \(10^{-6}\), except for specific humidity, for which \(10^{-10}\) was used. Initial and supervised residuals shared the same residual statistics. The same statistics were used at each 2~h residual update. Because standardization preceded rearrangement, all tiles of a given variable and level used the same statistics. These transformations remained fixed during validation and testing.

Training used 6~h windows, with the correction module unrolled for three shared-parameter updates at 2, 4 and 6~h. The initial residual was derived from the analysis at \(t_0\). Each later update received the preceding predicted residual and the next parent increment. The HRES-0.07\textdegree{} analysis at \(t_0+6\)~h supplied terminal supervision; intermediate residual predictions were propagated without replacement by reference residuals.

For terminal residual predictions \(\widehat y_i\) and targets \(y_i\) in standardized coordinates, the loss was

\begin{equation}
\mathcal L(\theta)
=
\frac{1}{N_{\mathrm e}}
\sum_{i=1}^{N_{\mathrm e}}
a_i^2(\widehat y_i-y_i)^2,
\qquad
a_i=\cos\phi_i,
\label{eq:supp_terminal_training_loss}
\end{equation}
where \(N_{\mathrm e}\) is the number of valid residual elements and \(\phi_i\) is latitude. Both fields are restored to the target-grid arrangement before loss evaluation, so the weights correspond to geographical latitude. This expression corresponds to applying \(\cos\phi_i\) to both predictions and targets before evaluating MSE, giving an effective squared-error weight of \(\cos^2\phi_i\). The same terminal loss was used for validation.

Gradients propagated through all three correction updates. Parent states and increments were fixed inputs to optimization, and only the correction-module parameters \(\theta\) were updated. Optimization used momentum SGD with a momentum coefficient of 0.9 and a mini-batch size of one. One optimizer update was performed per 6 h training window. The learning rate was adjusted in stages according to the observed progression of the training loss. Training started with a learning rate of \(10^{-3}\). Following the initial rapid decrease in training loss, the learning rate was reduced to \(5\times10^{-4}\) when the loss reached a plateau, approximately after the first epoch. Training continued at this learning rate for approximately 10 additional epochs, after which a further plateau prompted a reduction to \(10^{-4}\). The learning rate remained constant between these adjustments. The checkpoint used for evaluation was selected according to the 2021 validation loss.

At inference, the same standardized residual representation and 2~h update were retained throughout the forecast. Predicted residuals were propagated autoregressively beyond the 6~h training horizon, while the parent continued its own integration. No reference residuals were supplied after initialization.

\subsection{Training resources}
\label{subsec:supp_training_resources}

Table~\ref{tab:supp_training_resources} summarizes this configuration alongside training resources reported for other global forecasting models. The One2N-0.07\textdegree{} entry describes the incremental training required after the pretrained parent was available. The table provides context for hardware requirements and training duration; the entries represent different datasets, numerical precisions, training objectives and stages.

\begin{table}[htbp]
\centering
\small
\setlength{\tabcolsep}{3pt}
\renewcommand{\arraystretch}{1.12}
\caption{Training hardware and wall-clock time for global forecasting models. Grid spacing is nominal; Aurora uses multiple pretraining resolutions. Intervals denote model prediction steps and are not applicable to the continuous NeuralGCM-0.7\textdegree{} solver. Pangu-Weather-0.25\textdegree{} time is per interval-specific model, and FuXi-0.25\textdegree{} fine-tuning time is per cascaded model. One2N-0.07\textdegree{} time includes correction-module optimization and online integration of the frozen parent.}
\label{tab:supp_training_resources}
\begin{tabular*}{\textwidth}{@{\extracolsep{\fill}}lccllrr@{}}
\hline
Model & \shortstack{Grid(\textdegree{})}
& \shortstack{Interval(h)}
& Stage & Device & Count & Time \\
\hline
Pangu-Weather\cite{bi2023accurate}
& 0.25 & 1, 3, 6, 24 & Train & V100 GPU & 192 & $\sim$16 d \\
GraphCast\cite{lam2023learning}
& 0.25 & 6 & Train & TPU v4 & 32 & $\sim$4 weeks \\
NeuralGCM\cite{kochkov2024neural}
& 0.7 & --- & Train & TPU v4 & 256 & $\sim$3 weeks \\
FuXi\cite{chen2023fuxi}
& 0.25 & 6 & Pretrain & A100 GPU & 8 & $\sim$30 h \\
FuXi & 0.25 & 6 & Fine-tune & A100 GPU & 8 & $\sim$2 d \\
Aurora\cite{bodnar2025foundation}
& Mixed & 6 & Pretrain & A100 GPU & 32 & $\sim$2.5 weeks \\
One2N & 0.07 & 2 & Refine & A800 GPU & 1 & $\sim$4 weeks \\
\hline
\end{tabular*}

\end{table}

\section{Verification metrics and aggregation}
\label{sec:supp_note5}

\subsection{Verification setup}
\label{subsec:supp_metric}

Gridded verification uses the remapping, reference fields and common valid-cell masks described in Supplementary Section~\ref{sec:supp_note1}. Metrics are calculated from fields in their original physical units. Within each comparison, models use the same verifying analysis and matched initializations.

At a fixed initialization, variable, pressure level and lead time, let \(\widehat{x}_i\) and \(x_i\) denote forecast and reference values at grid cell \(i\), and let \(e_i=\widehat{x}_i-x_i\). Spatial metrics use latitude weights normalized over the common, non-empty valid-cell set \(\Omega\),

\begin{equation}
w_i
=
\frac{\cos\phi_i}
{\displaystyle\sum_{j\in\Omega}\cos\phi_j},
\qquad
\sum_{i\in\Omega}w_i=1,
\label{eq:latitude_weight}
\end{equation}
where \(\phi_i\) is latitude. Initialization, variable, level and lead-time indices are omitted from the spatial definitions below.

\subsection{Metric definitions}
\label{subsec:supp_metric_definitions}

\subsubsection{RMSE, MAE and bias}
\label{subsec:rmse}
\label{subsec:diagnostic_metrics}

The root-mean-square error, mean absolute error and signed bias are

\begin{align}
\mathrm{RMSE}
&=
\left(\sum_{i\in\Omega}w_i e_i^2\right)^{1/2},
\label{eq:rmse}\\
\mathrm{MAE}
&=
\sum_{i\in\Omega}w_i|e_i|,
\label{eq:supp_mae}\\
\mathrm{Bias}
&=
\sum_{i\in\Omega}w_i e_i.
\label{eq:supp_bias}
\end{align}
All three metrics have the units of the evaluated variable. Positive bias indicates overestimation relative to the reference.

\subsubsection{Normalized RMSE}

Normalized RMSE scales the error by the weighted spatial standard deviation of the reference field,

\begin{equation}
\mathrm{nRMSE}=
\frac{\mathrm{RMSE}}{\sigma_x},\quad
\sigma_x=
\left[\sum_{i\in\Omega}w_i(x_i-\bar{x})^2\right]^{1/2},
\quad
\bar{x}=\sum_{i\in\Omega}w_i x_i.
\label{eq:supp_nrmse}
\end{equation}
nRMSE is dimensionless, with the reference standard deviation computed separately for each initialization, variable, level and lead time. The same definitions are used for analysis-to-analysis diagnostics, with the second named analysis treated as the reference. For upper-stratospheric specific humidity, nRMSE is assessed alongside errors in physical units and spatial correlation because the reference spatial variability is small.

\subsubsection{Spatial correlation}

The weighted spatial correlation is

\begin{equation}
r=\frac{
\displaystyle\sum_{i\in\Omega}
w_i(\widehat{x}_i-\bar{\widehat{x}})(x_i-\bar{x})
}{
\displaystyle
\left[
\left(\sum_{i\in\Omega}w_i(\widehat{x}_i-\bar{\widehat{x}})^2\right)
\left(\sum_{i\in\Omega}w_i(x_i-\bar{x})^2\right)
\right]^{1/2}
},
\label{eq:spatial_correlation}
\end{equation}
where \(\bar{\widehat{x}}=\sum_{i\in\Omega}w_i\widehat{x}_i\) and \(\bar{x}\) is defined in Eq.~\eqref{eq:supp_nrmse}.

\subsubsection{Anomaly correlation coefficient}
\label{subsec:acc}

The anomaly correlation coefficient (ACC) is calculated relative to a climatological mean \(c_i\). The climatology is constructed from HRES-0.07\textdegree{} analyses for 2016--2020 using a \(\pm30\) day calendar window centred on the verification day of year and the matching synoptic hour (0000, 0600, 1200 or 1800 UTC). Defining \(a_i=x_i-c_i\) and \(\widehat{a}_i=\widehat{x}_i-c_i\) gives
\begin{equation}
\mathrm{ACC}
=
\frac{
\displaystyle\sum_{i\in\Omega}w_i\widehat{a}_i a_i
}{
\displaystyle
\left[
\left(\sum_{i\in\Omega}w_i\widehat{a}_i^2\right)
\left(\sum_{i\in\Omega}w_i a_i^2\right)
\right]^{1/2}
}.
\label{eq:acc}
\end{equation}
All models within a comparison use the same climatology on the verification grid. ACC is not reported for the upper-stratospheric diagnostics. Normalized errors and correlation coefficients are undefined when their denominators are zero.

\subsection{Aggregation and model comparison}
\label{subsec:aggregation}

\subsubsection{Averaging over initializations, lead times and levels}

Each spatial metric is first calculated separately for each initialization, variable, pressure level and lead time. For a given variable, let \(S_{m,\tau,\ell}\) denote a metric for initialization \(m\), lead time \(\tau\) and pressure level \(\ell\). The mean over \(M\) matched initializations is
\begin{equation}
\bar S_{\tau,\ell}
=
\frac{1}{M}\sum_{m=1}^{M}S_{m,\tau,\ell}.
\label{eq:sample_mean}
\end{equation}
For RMSE, the square root is therefore taken separately for each initialization before averaging.

The 10-day mean gives equal weight to the 20 verified lead times,
\begin{equation}
\langle S\rangle_{\ell}^{12:240}
=
\frac{1}{|\mathcal T|}
\sum_{\tau\in\mathcal T}\bar S_{\tau,\ell},
\qquad
\mathcal T=\{12,24,\ldots,240\}\ \mathrm{h}.
\label{eq:window_mean}
\end{equation}
Lead time zero is excluded. The day-10 score refers specifically to 240~h.

For the selected pressure levels \(\mathcal P\), expressed in hPa, the pressure bands are
\begin{align*}
\mathcal L_1&=\{\ell\in\mathcal P:1\leq\ell\leq30\},&
\mathcal L_2&=\{\ell\in\mathcal P:30<\ell\leq100\},\\
\mathcal L_3&=\{\ell\in\mathcal P:100<\ell\leq300\},&
\mathcal L_4&=\{\ell\in\mathcal P:300<\ell\leq1000\}.
\end{align*}
The pressure-band mean is
\begin{equation}
\langle S\rangle_b^{12:240}
=
\frac{1}{|\mathcal L_b|}
\sum_{\ell\in\mathcal L_b}
\langle S\rangle_{\ell}^{12:240},
\label{eq:band_mean}
\end{equation}
with equal weight assigned to each selected level. Level--lead-time scorecards retain each variable--level--lead-time combination after averaging over initializations. Mean bias maps are obtained by averaging signed errors over initializations at each grid cell.

\subsubsection{Relative RMSE reduction}
\label{subsec:skill_score}

For model \(A\) relative to baseline \(B\), the RMSE skill score is calculated from the initialization-mean RMSE values at each lead time and pressure level, provided that the baseline RMSE is positive,
\begin{equation}
S_{\mathrm{RMSE},\tau,\ell}(A\mid B)
=
1-
\frac{\overline{\mathrm{RMSE}}^{\,A}_{\tau,\ell}}
     {\overline{\mathrm{RMSE}}^{\,B}_{\tau,\ell}}.
\label{eq:rmse_skill}
\end{equation}
Here, both models are evaluated against the same analysis using the same grid, mask and matched initializations. Positive skill indicates lower RMSE for \(A\), and the percentage reduction is \(100S_{\mathrm{RMSE},\tau,\ell}\).

The lead-time-averaged percentage reduction is
\begin{equation}
\overline{\Delta}_{\mathrm{RMSE},\ell}^{12:240}(A\mid B)
=
\frac{100}{|\mathcal T|}
\sum_{\tau\in\mathcal T}
S_{\mathrm{RMSE},\tau,\ell}(A\mid B).
\label{eq:supp_mean_relative_reduction}
\end{equation}
Thus, relative reductions are calculated at each lead time before averaging. Pressure-band summaries of these reductions give equal weight to the selected levels.

For the fraction of targets with lower RMSE, each variable--pressure-level--lead-time combination is counted once. A target is classified as improved when the initialization-mean RMSE of \(A\) is strictly lower than that of \(B\); the improved-target count is divided by the total number of evaluated combinations.

The global RMSE and ACC curves and the level--lead-time skill scorecards are reported as point estimates. Tropical-cyclone verification and its confidence intervals are described in Supplementary Section~\ref{sec:supp_note7}.

\section{Additional global forecast evaluation}
\label{sec:supp_forecast_results}

We extend the main-text evaluation to additional variables and pressure levels and repeat the comparisons against ERA5-0.25\textdegree{}. HRES-0.07\textdegree{}-based and ERA5-0.25\textdegree{}-based scores are computed on common 0.07\textdegree{} and 0.25\textdegree{} grids, respectively, using the verification procedures in Supplementary Section~\ref{sec:supp_note5}. Forecasts are evaluated every 12~h from 12 to 240~h. Comparisons with NeuralGCM-0.7\textdegree{} cover five variables at 25 pressure levels; those with Pangu-Weather-0.25\textdegree{} use 13 shared levels from 1000 to 50~hPa.

Z, T, U, V and Q denote geopotential, temperature, zonal wind, meridional wind and specific humidity, respectively. Skill scores are defined in Eq.~\ref{eq:rmse_skill}; positive values indicate lower RMSE for One2N-0.07\textdegree{}. Tables report relative RMSE reductions at 240~h, expressed as percentages, rather than averages over forecast lead times. ACC is reported only for the tropospheric comparisons.

\subsection{Tropospheric forecast skill against HRES-0.07\textdegree{}}
\label{subsec:s1_tropospheric}

\begin{figure}[t]
\centering
\includegraphics[width=\linewidth,height=0.70\textheight,keepaspectratio]{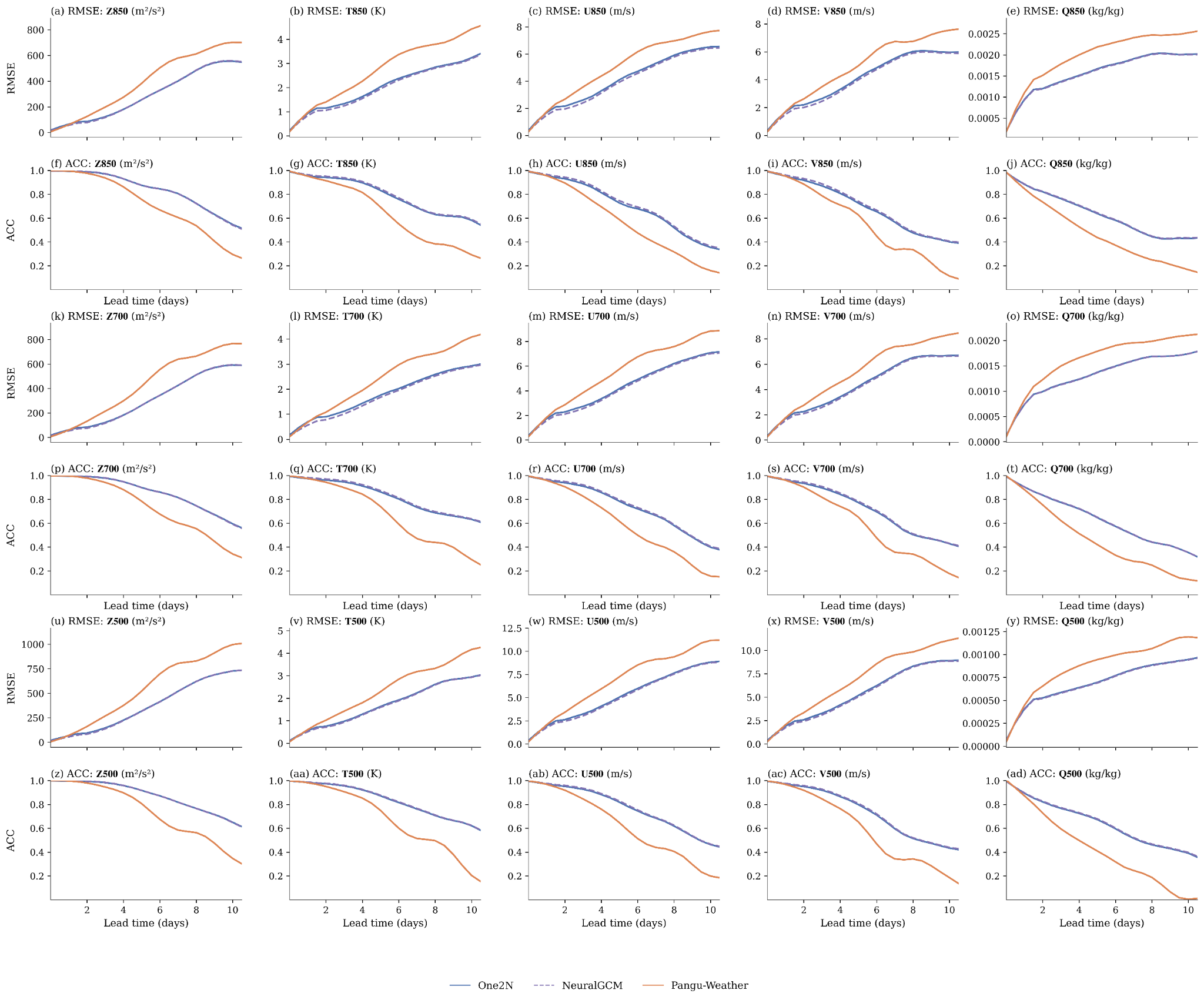}
\caption[Forecast skill at representative tropospheric levels]{\textbf{Forecast skill at representative tropospheric levels.} Latitude-weighted RMSE and ACC for One2N-0.07\textdegree{}, NeuralGCM-0.7\textdegree{} and Pangu-Weather-0.25\textdegree{} at 850, 700 and 500~hPa, evaluated against HRES-0.07\textdegree{} analyses on the common 0.07\textdegree{} grid. Results are shown for Z, T, U, V and Q. Lower RMSE and higher ACC indicate greater skill.}
\label{sfig:forecast_skill_five}
\end{figure}

At 850, 700 and 500~hPa, One2N-0.07\textdegree{} has broadly comparable RMSE and ACC to NeuralGCM-0.7\textdegree{} (Supplementary Fig.~\ref{sfig:forecast_skill_five}). Day-10 relative RMSE reductions range from \(-2.52\%\) to \(+0.56\%\) across the 15 variable--level combinations (Supplementary Table~\ref{tab:s1_tropo_comparison}).

Compared with Pangu-Weather-0.25\textdegree{}, One2N-0.07\textdegree{} has lower day-10 RMSE for all 15 combinations, with reductions of \(16.38\)--\(30.61\%\). Across these levels and all 20 lead times, One2N-0.07\textdegree{} has lower RMSE for 283 of 300 targets (\(94.33\%\)).

\begin{table*}[t]
\centering
\small
\renewcommand{\arraystretch}{1.12}
\setlength{\tabcolsep}{9pt}
\caption[Day-10 relative RMSE reductions at representative tropospheric levels]{\textbf{Day-10 relative RMSE reductions at representative tropospheric levels.} One2N-0.07\textdegree{} is compared with (a) NeuralGCM-0.7\textdegree{} and (b) Pangu-Weather-0.25\textdegree{}, using HRES-0.07\textdegree{} analyses on the common 0.07\textdegree{} grid. Values are percentages; positive values indicate lower RMSE for One2N-0.07\textdegree{}.}
\label{tab:s1_tropo_comparison}
\begin{tabular}{rrrrrr}
\multicolumn{6}{c}{\textbf{(a) One2N-0.07\textdegree{} relative to NeuralGCM-0.7\textdegree{}}}\\
\hline
Pressure (hPa) & Z & T & U & V & Q\\
\hline
850 & \(+0.56\) & \(-2.01\) & \(-2.52\) & \(-2.11\) & \(-0.90\)\\
700 & \(+0.32\) & \(-1.89\) & \(-1.96\) & \(-1.44\) & \(-0.41\)\\
500 & \(-0.01\) & \(-0.92\) & \(-1.52\) & \(-1.40\) & \(-0.69\)\\
\hline
\multicolumn{6}{c}{}\\[-2mm]
\multicolumn{6}{c}{\textbf{(b) One2N-0.07\textdegree{} relative to Pangu-Weather-0.25\textdegree{}}}\\
\hline
Pressure (hPa) & Z & T & U & V & Q\\
\hline
850 & \(20.66\) & \(27.17\) & \(16.38\) & \(19.26\) & \(20.21\)\\
700 & \(22.83\) & \(28.38\) & \(20.35\) & \(18.09\) & \(16.40\)\\
500 & \(27.71\) & \(30.61\) & \(21.58\) & \(18.35\) & \(20.17\)\\
\hline
\end{tabular}
\vspace{2mm}

\end{table*}

\subsection{Vertical structure of forecast skill against HRES-0.07\textdegree{}}
\label{subsec:s1_vertical}

\begin{figure}[t]
\centering
\includegraphics[width=\linewidth,height=0.70\textheight,keepaspectratio]{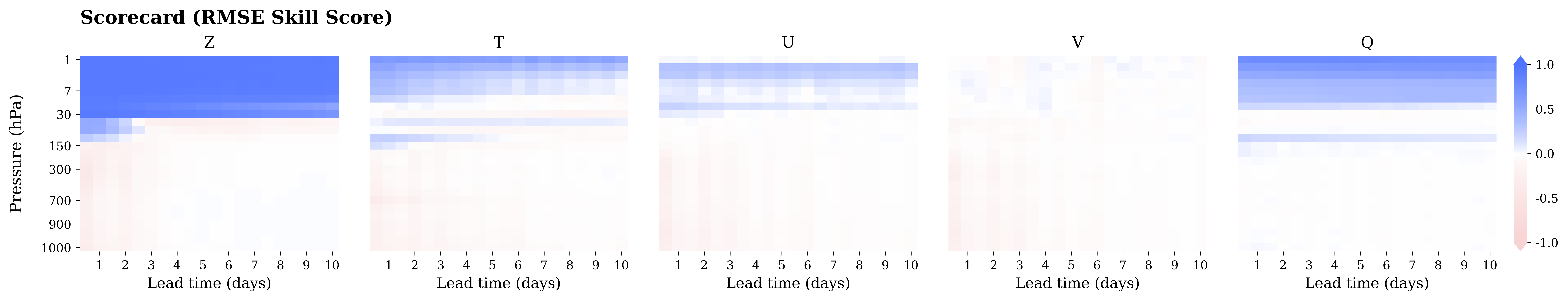}
\caption[RMSE skill relative to NeuralGCM-0.7\textdegree{} under HRES-0.07\textdegree{} verification]{\textbf{RMSE skill relative to NeuralGCM-0.7\textdegree{} under HRES-0.07\textdegree{} verification.} Skill scores for One2N-0.07\textdegree{} across five variables, 25 pressure levels and lead times from 12 to 240~h, evaluated against HRES-0.07\textdegree{} analyses on the common 0.07\textdegree{} grid. Positive values indicate lower RMSE for One2N-0.07\textdegree{}.}
\label{sfig:vertical_scorecard}
\end{figure}

The full-column comparison shows that the largest improvements are concentrated in upper-stratospheric geopotential, while lower- and middle-tropospheric differences are generally modest (Supplementary Fig.~\ref{sfig:vertical_scorecard}). At day~10, geopotential RMSE is reduced by \(88.39\)--\(90.97\%\) across 1--7~hPa, and temperature RMSE at 1~hPa is reduced by \(50.99\%\) (Supplementary Table~\ref{tab:s1_vertical}).

Changes in wind and humidity are more variable across levels. Some stratospheric metrics deteriorate, including geopotential at 50~hPa and temperature at 30~hPa. The largest upper-level gains therefore coexist with local increases in error.

\begin{table*}[t]
\centering
\small
\renewcommand{\arraystretch}{1.12}
\setlength{\tabcolsep}{9pt}
\caption[Day-10 relative RMSE reductions across 25 pressure levels under HRES-0.07\textdegree{} verification]{\textbf{Day-10 relative RMSE reductions across 25 pressure levels under HRES-0.07\textdegree{} verification.} One2N-0.07\textdegree{} is compared with NeuralGCM-0.7\textdegree{}, using HRES-0.07\textdegree{} analyses on the common 0.07\textdegree{} grid. Values are percentages; positive values indicate lower RMSE for One2N-0.07\textdegree{}.}
\label{tab:s1_vertical}
\begin{tabular}{rrrrrr}
\hline
Pressure (hPa) & Z & T & U & V & Q \\
\hline
1000 & \(+0.48\) & \(-4.22\) & \(-3.55\) & \(-2.46\) & \(-0.47\) \\
950 & \(+0.48\) & \(-3.21\) & \(-3.27\) & \(-2.31\) & \(-0.78\) \\
925 & \(+0.54\) & \(-2.70\) & \(-3.17\) & \(-2.30\) & \(-1.03\) \\
900 & \(+0.58\) & \(-2.47\) & \(-2.96\) & \(-2.29\) & \(-1.14\) \\
850 & \(+0.56\) & \(-2.01\) & \(-2.52\) & \(-2.11\) & \(-0.90\) \\
800 & \(+0.52\) & \(-2.12\) & \(-2.26\) & \(-1.95\) & \(-0.72\) \\
700 & \(+0.32\) & \(-1.89\) & \(-1.96\) & \(-1.44\) & \(-0.41\) \\
600 & \(+0.13\) & \(-1.32\) & \(-1.60\) & \(-1.29\) & \(-0.62\) \\
500 & \(-0.01\) & \(-0.92\) & \(-1.52\) & \(-1.40\) & \(-0.69\) \\
400 & \(-0.14\) & \(-0.75\) & \(-1.57\) & \(-1.52\) & \(-0.66\) \\
300 & \(-0.23\) & \(-1.11\) & \(-1.72\) & \(-1.45\) & \(-0.65\) \\
250 & \(-0.19\) & \(-0.88\) & \(-1.54\) & \(-1.18\) & \(-0.49\) \\
200 & \(-0.12\) & \(-0.95\) & \(-1.38\) & \(-1.03\) & \(+0.10\) \\
150 & \(-0.29\) & \(-3.10\) & \(-1.23\) & \(-0.83\) & \(-0.67\) \\
100 & \(-1.69\) & \(-6.79\) & \(-0.59\) & \(-0.77\) & \(+8.12\) \\
70 & \(-7.79\) & \(-4.83\) & \(-1.29\) & \(-1.67\) & \(+0.09\) \\
50 & \(-12.44\) & \(+5.39\) & \(-1.06\) & \(-1.73\) & \(-0.75\) \\
30 & \(+69.41\) & \(-13.65\) & \(-3.74\) & \(-3.01\) & \(-1.43\) \\
20 & \(+55.26\) & \(-3.83\) & \(+3.95\) & \(-1.45\) & \(+2.75\) \\
10 & \(+81.93\) & \(-6.03\) & \(-0.68\) & \(-1.66\) & \(+37.05\) \\
7 & \(+88.39\) & \(+1.65\) & \(+0.45\) & \(-0.82\) & \(+38.46\) \\
5 & \(+90.15\) & \(+1.21\) & \(+6.76\) & \(-2.47\) & \(+42.70\) \\
3 & \(+90.63\) & \(+10.96\) & \(+20.48\) & \(-2.57\) & \(+57.62\) \\
2 & \(+90.61\) & \(+30.39\) & \(+26.92\) & \(-3.86\) & \(+66.59\) \\
1 & \(+90.97\) & \(+50.99\) & \(-1.08\) & \(-5.15\) & \(+73.97\) \\
\hline
\end{tabular}
\vspace{2mm}

\end{table*}

\subsection{Full comparison with Pangu-Weather-0.25\textdegree{} against HRES-0.07\textdegree{}}
\label{subsec:hres_pangu_full}

\begin{figure}[htbp]
\centering
\includegraphics[width=\linewidth,height=0.70\textheight,keepaspectratio]{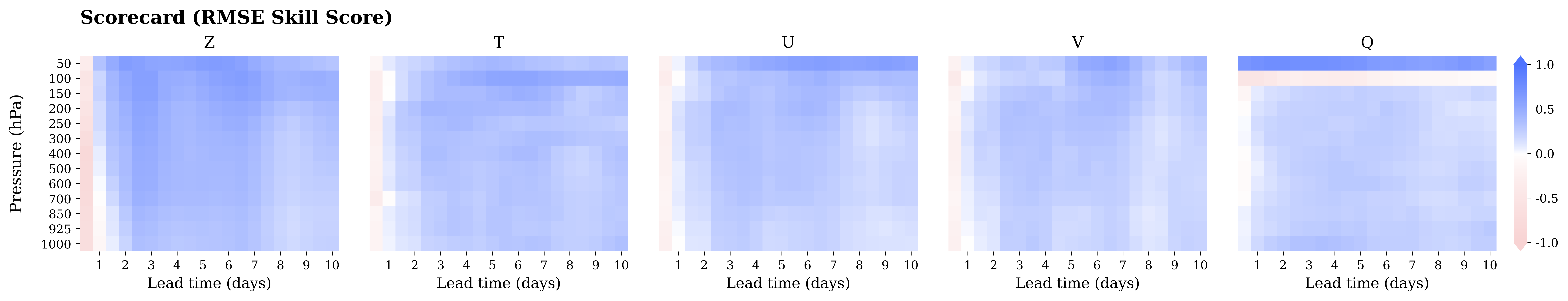}
\caption[RMSE skill relative to Pangu-Weather-0.25\textdegree{} under HRES-0.07\textdegree{} verification]{\textbf{RMSE skill relative to Pangu-Weather-0.25\textdegree{} under HRES-0.07\textdegree{} verification.} One2N-0.07\textdegree{} is compared with the repeated-3~h Pangu-Weather-0.25\textdegree{} configuration for five variables on the 13 shared pressure levels and lead times from 12 to 240~h. Forecasts are evaluated against HRES-0.07\textdegree{} on the common 0.07\textdegree{} grid. The score is defined in Eq.~\ref{eq:rmse_skill}; positive values indicate lower RMSE for One2N-0.07\textdegree{}.}
\label{sfig:pangu_hres_full}
\end{figure}

Across five variables, the 13 shared pressure levels and 20 non-zero lead times, One2N-0.07\textdegree{} has lower initialization-mean RMSE than the repeated-3~h Pangu-Weather-0.25\textdegree{} configuration for 1,215 of 1,300 targets (93.46\%; Supplementary Fig.~\ref{sfig:pangu_hres_full}). This target count assigns equal weight to every variable--level--lead-time combination and is a descriptive summary of point estimates.

\subsection{Evaluation against ERA5-0.25\textdegree{}}
\label{subsec:era5_robustness}

We repeat the evaluation against ERA5-0.25\textdegree{} on the common 0.25\textdegree{} grid, retaining the same forecast trajectories and initializations. This evaluation changes both the reference analysis and the verification grid.

\subsubsection{Comparison with NeuralGCM-0.7\textdegree{}}
\label{subsec:s2_neural}

\begin{figure}[t]
\centering
\includegraphics[width=\linewidth,height=0.70\textheight,keepaspectratio]{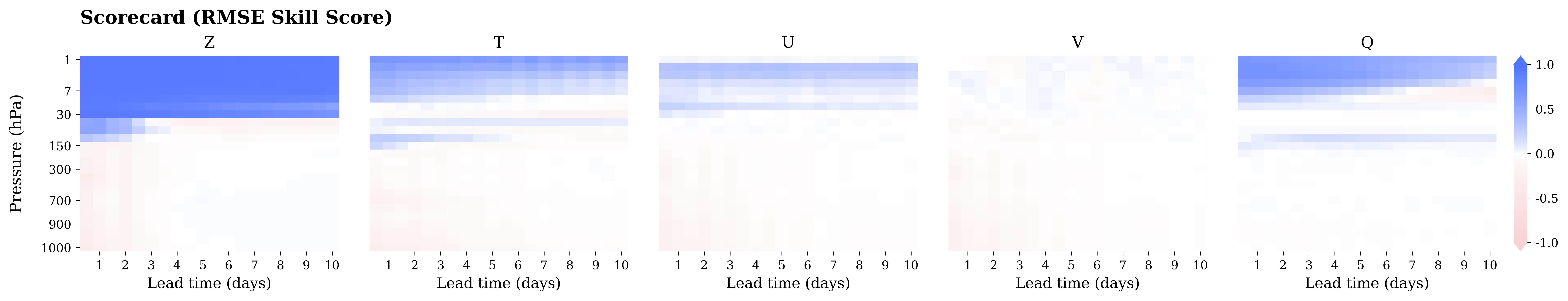}
\caption[RMSE skill relative to NeuralGCM-0.7\textdegree{} under ERA5-0.25\textdegree{} verification]{\textbf{RMSE skill relative to NeuralGCM-0.7\textdegree{} under ERA5-0.25\textdegree{} verification.} Skill scores for One2N-0.07\textdegree{} across five variables, 25 pressure levels and lead times from 12 to 240~h, evaluated against ERA5-0.25\textdegree{} analyses on the common 0.25\textdegree{} grid. Positive values indicate lower RMSE for One2N-0.07\textdegree{}.}
\label{sfig:vertical_scorecard_neuralgcm_0.25}
\end{figure}

The ERA5-0.25\textdegree{}-based evaluation shows a similar vertical pattern (Supplementary Fig.~\ref{sfig:vertical_scorecard_neuralgcm_0.25}). At day~10, RMSE reductions at 850, 700 and 500~hPa range from \(-1.59\%\) to \(+0.67\%\), while geopotential reductions across 1--7~hPa range from \(88.40\%\) to \(91.33\%\). Temperature RMSE at 1~hPa is reduced by \(55.83\%\) (Supplementary Table~\ref{tab:s2_neural_era5}).

Upper-level humidity is more sensitive to the verification configuration: day-10 RMSE increases at 10, 7 and 5~hPa under ERA5-0.25\textdegree{} verification, despite decreasing against HRES-0.07\textdegree{}. The large geopotential reductions and selected temperature improvements are more consistent across the two references. Bias and analysis-reference diagnostics are presented in Supplementary Section~\ref{sec:one2n_mechanism}.

\begin{table*}[t]
\centering
\small
\renewcommand{\arraystretch}{1.12}
\setlength{\tabcolsep}{9pt}
\caption[Day-10 relative RMSE reductions across 25 pressure levels under ERA5-0.25\textdegree{} verification]{\textbf{Day-10 relative RMSE reductions across 25 pressure levels under ERA5-0.25\textdegree{} verification.} One2N-0.07\textdegree{} is compared with NeuralGCM-0.7\textdegree{}, using ERA5-0.25\textdegree{} analyses on the common 0.25\textdegree{} grid. Values are percentages; positive values indicate lower RMSE for One2N-0.07\textdegree{}.}
\label{tab:s2_neural_era5}
\begin{tabular}{rrrrrr}
\hline
Pressure (hPa) & Z & T & U & V & Q \\
\hline
1000 & \(+0.65\) & \(-3.73\) & \(-2.61\) & \(-1.52\) & \(-0.54\) \\
950 & \(+0.64\) & \(-2.60\) & \(-2.30\) & \(-1.27\) & \(-0.76\) \\
925 & \(+0.68\) & \(-2.08\) & \(-2.25\) & \(-1.22\) & \(-1.12\) \\
900 & \(+0.68\) & \(-1.74\) & \(-2.00\) & \(-1.23\) & \(-1.04\) \\
850 & \(+0.67\) & \(-1.41\) & \(-1.59\) & \(-1.13\) & \(-0.50\) \\
800 & \(+0.65\) & \(-1.54\) & \(-1.42\) & \(-1.05\) & \(-0.14\) \\
700 & \(+0.46\) & \(-1.29\) & \(-1.26\) & \(-0.74\) & \(-0.07\) \\
600 & \(+0.27\) & \(-0.80\) & \(-0.91\) & \(-0.64\) & \(-0.36\) \\
500 & \(+0.09\) & \(-0.46\) & \(-0.77\) & \(-0.74\) & \(-0.42\) \\
400 & \(+0.03\) & \(-0.30\) & \(-0.77\) & \(-0.76\) & \(-0.32\) \\
300 & \(-0.06\) & \(-0.49\) & \(-0.93\) & \(-0.80\) & \(-0.31\) \\
250 & \(-0.05\) & \(-0.27\) & \(-0.81\) & \(-0.61\) & \(-0.26\) \\
200 & \(+0.01\) & \(-0.51\) & \(-0.78\) & \(-0.61\) & \(+0.11\) \\
150 & \(-0.19\) & \(-2.49\) & \(-0.68\) & \(-0.49\) & \(+1.27\) \\
100 & \(-1.42\) & \(-5.61\) & \(-0.10\) & \(-0.19\) & \(+7.95\) \\
70 & \(-7.11\) & \(-4.05\) & \(-0.76\) & \(-0.89\) & \(+0.51\) \\
50 & \(-12.03\) & \(+5.87\) & \(-0.54\) & \(-1.11\) & \(-0.53\) \\
30 & \(+70.52\) & \(-12.21\) & \(-2.88\) & \(-2.14\) & \(-0.72\) \\
20 & \(+56.74\) & \(-3.26\) & \(+5.03\) & \(-0.74\) & \(+0.59\) \\
10 & \(+81.91\) & \(-2.58\) & \(+0.84\) & \(-0.67\) & \(-18.85\) \\
7 & \(+88.40\) & \(+7.17\) & \(+3.93\) & \(+0.45\) & \(-29.01\) \\
5 & \(+90.22\) & \(+9.52\) & \(+10.15\) & \(-0.83\) & \(-3.66\) \\
3 & \(+90.82\) & \(+22.66\) & \(+22.90\) & \(-0.51\) & \(+22.61\) \\
2 & \(+90.92\) & \(+40.87\) & \(+28.67\) & \(-1.28\) & \(+26.06\) \\
1 & \(+91.33\) & \(+55.83\) & \(+2.38\) & \(-2.68\) & \(+35.51\) \\
\hline
\end{tabular}
\vspace{2mm}

\end{table*}

\subsubsection{Comparison with Pangu-Weather-0.25\textdegree{}}
\label{subsec:s2_pangu}

\begin{figure}[t]
\centering
\includegraphics[width=\linewidth,height=0.70\textheight,keepaspectratio]{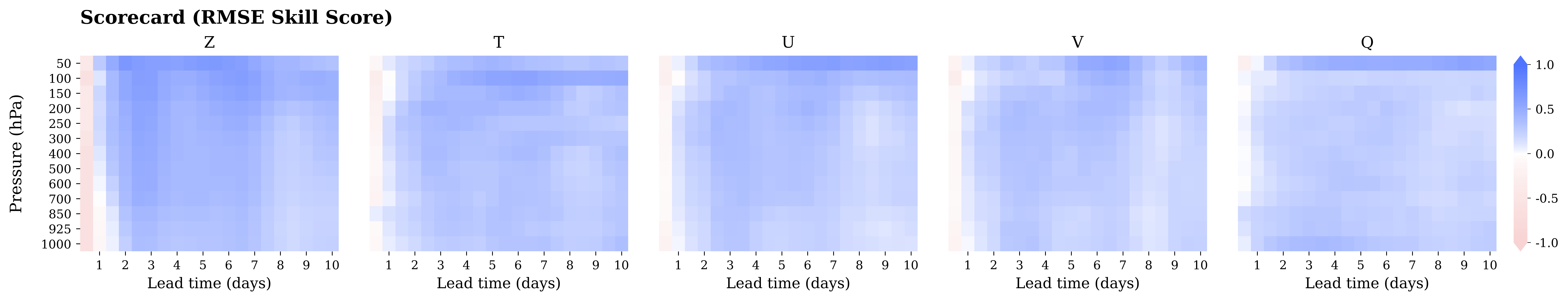}
\caption[RMSE skill relative to Pangu-Weather-0.25\textdegree{} under ERA5-0.25\textdegree{} verification]{\textbf{RMSE skill relative to Pangu-Weather-0.25\textdegree{} under ERA5-0.25\textdegree{} verification.} Skill scores for One2N-0.07\textdegree{} across five variables, the 13 shared pressure levels from 1000 to 50~hPa, and lead times from 12 to 240~h. Forecasts are evaluated against ERA5-0.25\textdegree{} analyses on the common 0.25\textdegree{} grid. Positive values indicate lower RMSE for One2N-0.07\textdegree{}.}
\label{sfig:vertical_scorecard_pangu_0.25}
\end{figure}

Against ERA5-0.25\textdegree{}, One2N-0.07\textdegree{} has lower RMSE than the evaluated Pangu-Weather-0.25\textdegree{} configuration for 1,239 of 1,300 common targets (\(95.31\%\); Supplementary Fig.~\ref{sfig:vertical_scorecard_pangu_0.25}). At day~10, RMSE is lower for all 65 variable--level combinations, with reductions of \(12.34\)--\(57.30\%\) (Supplementary Table~\ref{tab:s2_pangu_era5}). The comparison includes 50~hPa, the uppermost shared level, and excludes 70~hPa.

\begin{table*}[t]
\centering
\small
\renewcommand{\arraystretch}{1.12}
\setlength{\tabcolsep}{9pt}
\caption[Day-10 relative RMSE reductions at the 13 shared pressure levels under ERA5-0.25\textdegree{} verification]{\textbf{Day-10 relative RMSE reductions at the 13 shared pressure levels under ERA5-0.25\textdegree{} verification.} One2N-0.07\textdegree{} is compared with Pangu-Weather-0.25\textdegree{}, using ERA5-0.25\textdegree{} analyses on the common 0.25\textdegree{} grid. Values are percentages; positive values indicate lower RMSE for One2N-0.07\textdegree{}.}
\label{tab:s2_pangu_era5}
\begin{tabular}{rrrrrr}
\hline
Pressure (hPa) & Z & T & U & V & Q \\
\hline
1000 & \(+22.70\) & \(+34.98\) & \(+12.34\) & \(+21.96\) & \(+26.03\) \\
925 & \(+21.26\) & \(+29.61\) & \(+14.17\) & \(+21.61\) & \(+25.83\) \\
850 & \(+20.63\) & \(+29.27\) & \(+17.02\) & \(+20.10\) & \(+20.02\) \\
700 & \(+22.70\) & \(+28.84\) & \(+21.18\) & \(+19.33\) & \(+17.30\) \\
600 & \(+25.04\) & \(+29.10\) & \(+21.07\) & \(+18.93\) & \(+22.21\) \\
500 & \(+27.71\) & \(+31.25\) & \(+22.44\) & \(+18.98\) & \(+20.41\) \\
400 & \(+30.46\) & \(+34.15\) & \(+21.86\) & \(+19.57\) & \(+17.10\) \\
300 & \(+32.90\) & \(+29.11\) & \(+22.04\) & \(+21.95\) & \(+15.07\) \\
250 & \(+35.37\) & \(+22.70\) & \(+23.85\) & \(+24.49\) & \(+14.26\) \\
200 & \(+38.65\) & \(+31.97\) & \(+27.89\) & \(+28.18\) & \(+13.97\) \\
150 & \(+44.87\) & \(+32.98\) & \(+33.02\) & \(+27.66\) & \(+19.60\) \\
100 & \(+45.10\) & \(+50.42\) & \(+37.24\) & \(+34.91\) & \(+19.46\) \\
50 & \(+32.35\) & \(+28.79\) & \(+57.30\) & \(+43.13\) & \(+53.11\) \\
\hline
\end{tabular}
\vspace{2mm}

\end{table*}

\section{Cross-reference diagnosis of vertically varying forecast improvements}
\label{sec:one2n_mechanism}

To examine the vertical contrast in One2N-0.07\textdegree{} improvements, we compare differences between HRES-0.07\textdegree{} and ERA5-0.25\textdegree{} with the forecast errors of NeuralGCM-0.7\textdegree{} against each analysis, and then evaluate One2N-0.07\textdegree{} against both references. The analysis-to-analysis comparison provides context for assessing whether the parent-model error depends strongly on the verification reference. All fields are matched by valid time and evaluated on the common 0.25\textdegree{} grid using the common valid-cell mask specified in Supplementary Section~\ref{subsec:verification_grids}. Results are averaged over 12--240~h following Eq.~\ref{eq:window_mean}; metrics and pressure-band aggregation follow Supplementary Section~\ref{sec:supp_note5}.

\subsection{Reference differences and parent-model errors}
\label{subsec:reference_parent_error}

\begin{figure}[htbp]
\centering
\includegraphics[width=\linewidth,height=0.70\textheight,keepaspectratio]{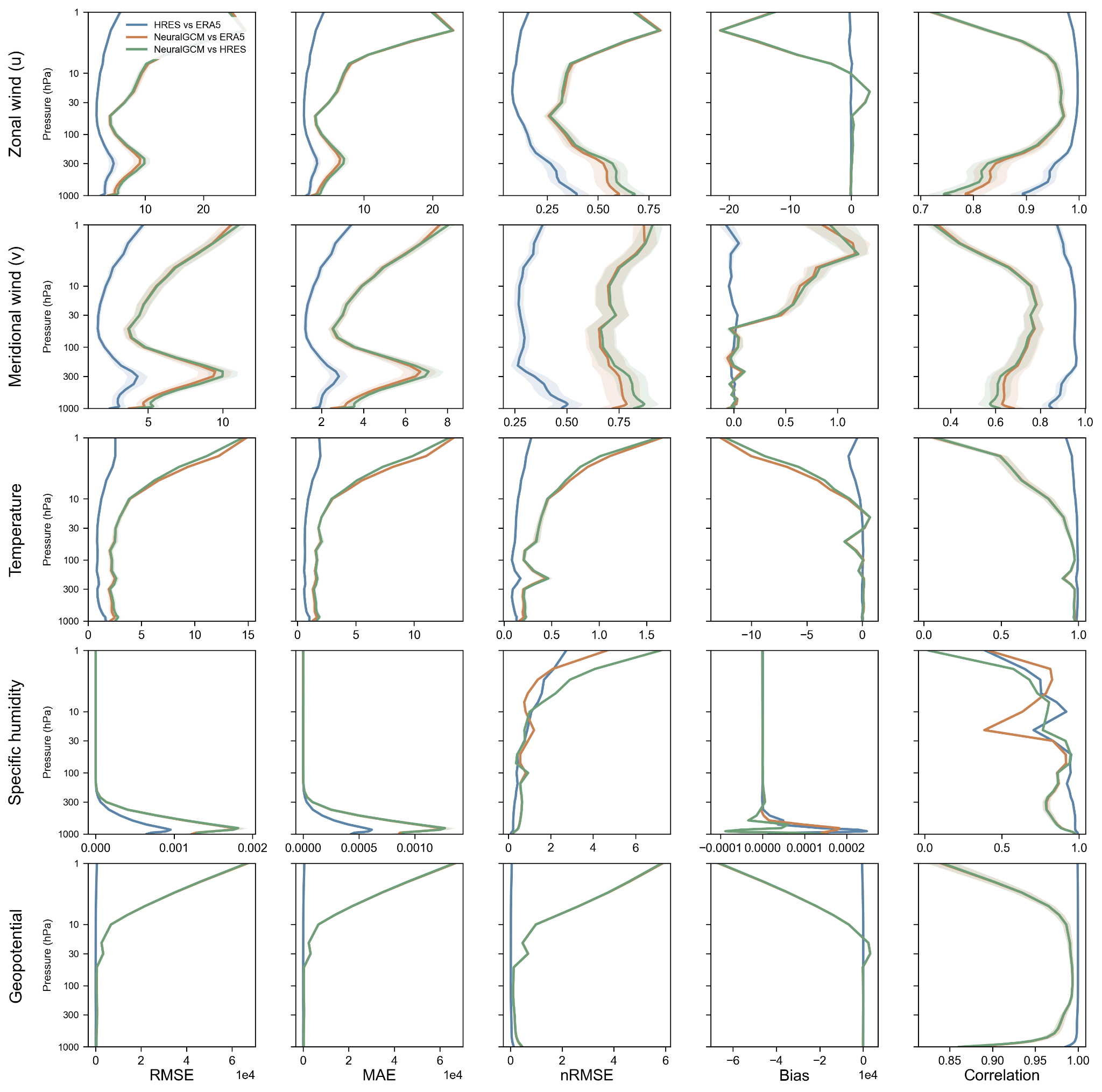}
\caption[Vertical profiles of reference differences and parent-model errors]{\textbf{Vertical profiles of reference differences and parent-model errors.}
RMSE, MAE, nRMSE, bias and spatial correlation for U, V, T, Q and Z at 25 pressure levels. Blue denotes HRES-0.07\textdegree{} versus ERA5-0.25\textdegree{}, orange NeuralGCM-0.7\textdegree{} versus ERA5-0.25\textdegree{}, and green NeuralGCM-0.7\textdegree{} versus HRES-0.07\textdegree{}. Lines show averages over matched initializations and lead times from 12 to 240~h; shading denotes 95\% confidence intervals. All comparisons use the common 0.25\textdegree{} grid.}
\label{fig:supp_vertical_diagnostics}
\end{figure}

The levelwise profiles distinguish changes in total error from changes in mean offset (Supplementary Fig.~\ref{fig:supp_vertical_diagnostics}). For the representative tropospheric metrics examined, NeuralGCM-0.7\textdegree{} has a small mean bias relative to its RMSE. At 1--7~hPa, geopotential and temperature instead exhibit large mean biases. For each variable, the bias has the same sign against both analyses, and its magnitude substantially exceeds the corresponding analysis-to-analysis difference. This agreement identifies a systematic component of the parent model’s upper-level error that is present under either reference.

\begin{figure}[htbp]
\centering
\includegraphics[width=\linewidth,height=0.70\textheight,keepaspectratio]{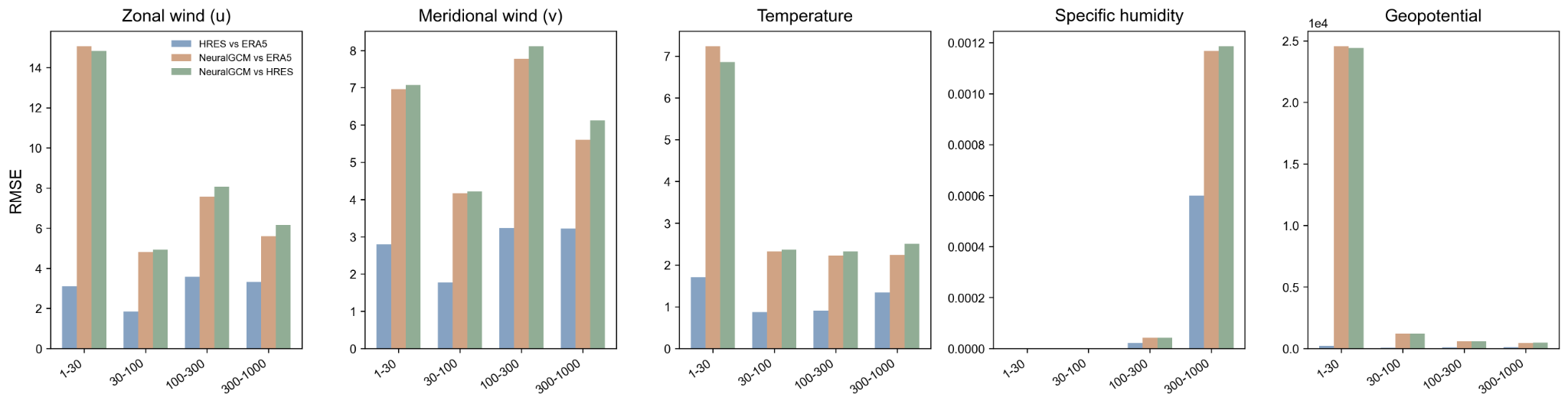}
\caption[Pressure-band RMSE of reference differences and parent-model errors]{\textbf{Pressure-band RMSE of reference differences and parent-model errors.}
Mean RMSE over 12--240~h for HRES-0.07\textdegree{} versus ERA5-0.25\textdegree{}, NeuralGCM-0.7\textdegree{} versus ERA5-0.25\textdegree{}, and NeuralGCM-0.7\textdegree{} versus HRES-0.07\textdegree{}. The four pressure bands follow Supplementary Section~\ref{subsec:aggregation}. All comparisons use the common 0.25\textdegree{} grid.}
\label{fig:supp_pressure_bar}
\end{figure}

Pressure-band RMSE places these offsets in the context of the broader error structure (Supplementary Fig.~\ref{fig:supp_pressure_bar}). In the 1--30-hPa band, geopotential RMSE is approximately \(2.44\times10^{4}\) and \(2.46\times10^{4}~\mathrm{m^2\,s^{-2}}\) for NeuralGCM-0.7\textdegree{} against HRES-0.07\textdegree{} and ERA5-0.25\textdegree{}, respectively, compared with \(219~\mathrm{m^2\,s^{-2}}\) between the analyses. The corresponding temperature RMSEs are 6.86, 7.24 and 1.71~K. The two parent-model error estimates are therefore similar and substantially larger than the reference discrepancy. The large upper-level errors cannot be explained by differences between HRES-0.07\textdegree{} and ERA5-0.25\textdegree{} alone. This separation is weaker for zonal wind and limited for meridional wind, indicating a variable-dependent error structure.

\subsection{One2N-0.07\textdegree{} improvements against both analyses}
\label{subsec:cross_reference_robustness}

\begin{figure}[htbp]
\centering
\includegraphics[width=\linewidth,height=0.70\textheight,keepaspectratio]{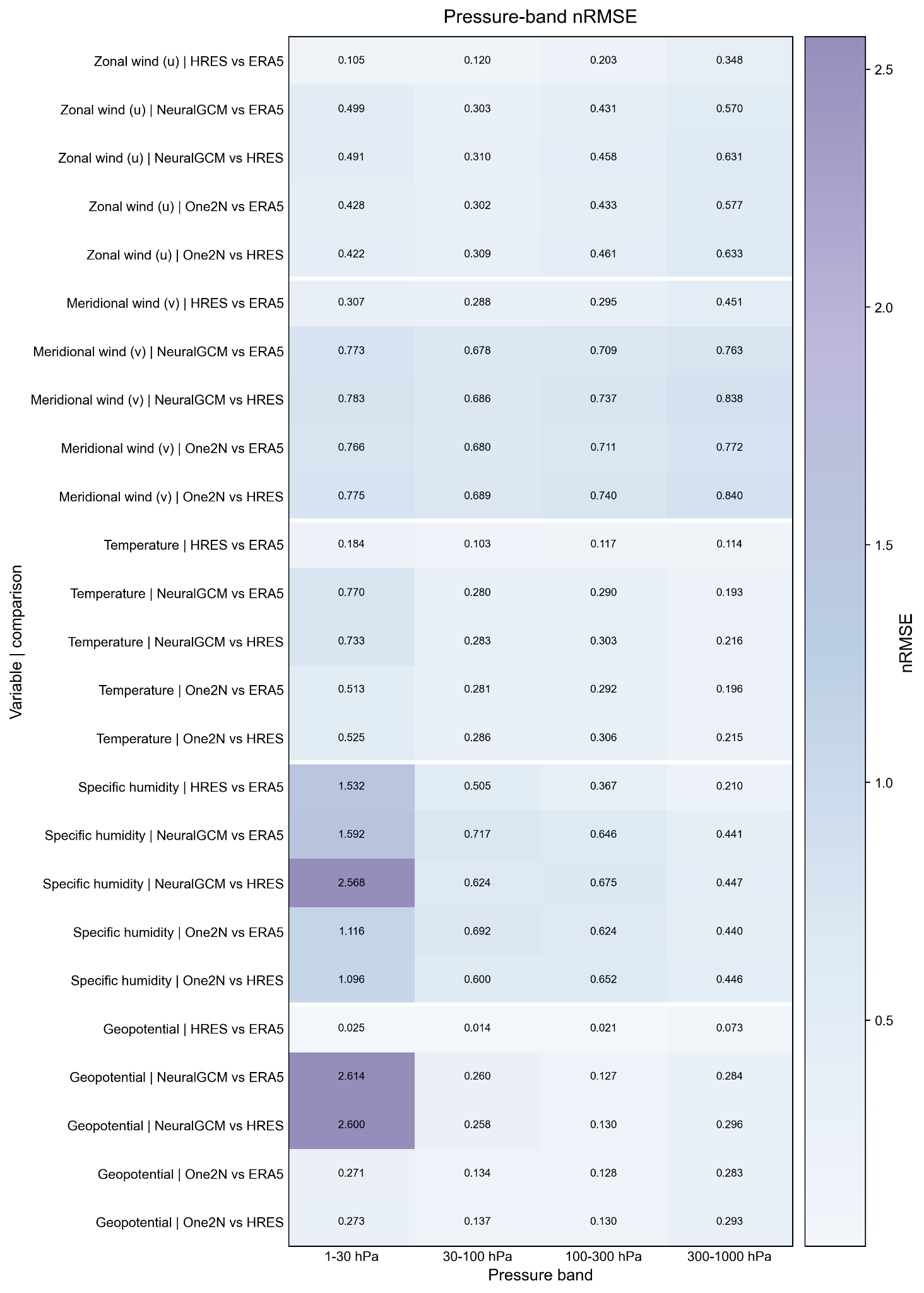}
\caption[Pressure-band nRMSE against alternative reference analyses]{\textbf{Pressure-band nRMSE against alternative reference analyses.}
Mean nRMSE over 12--240~h for HRES-0.07\textdegree{} versus ERA5-0.25\textdegree{}, and for NeuralGCM-0.7\textdegree{} and One2N-0.07\textdegree{} against each analysis. Pressure bands follow Supplementary Section~\ref{subsec:aggregation}. All comparisons use the common 0.25\textdegree{} grid.}
\label{fig:supp_pressure_nrmse}
\end{figure}

\begin{figure}[htbp]
\centering
\includegraphics[width=\linewidth,height=0.70\textheight,keepaspectratio]{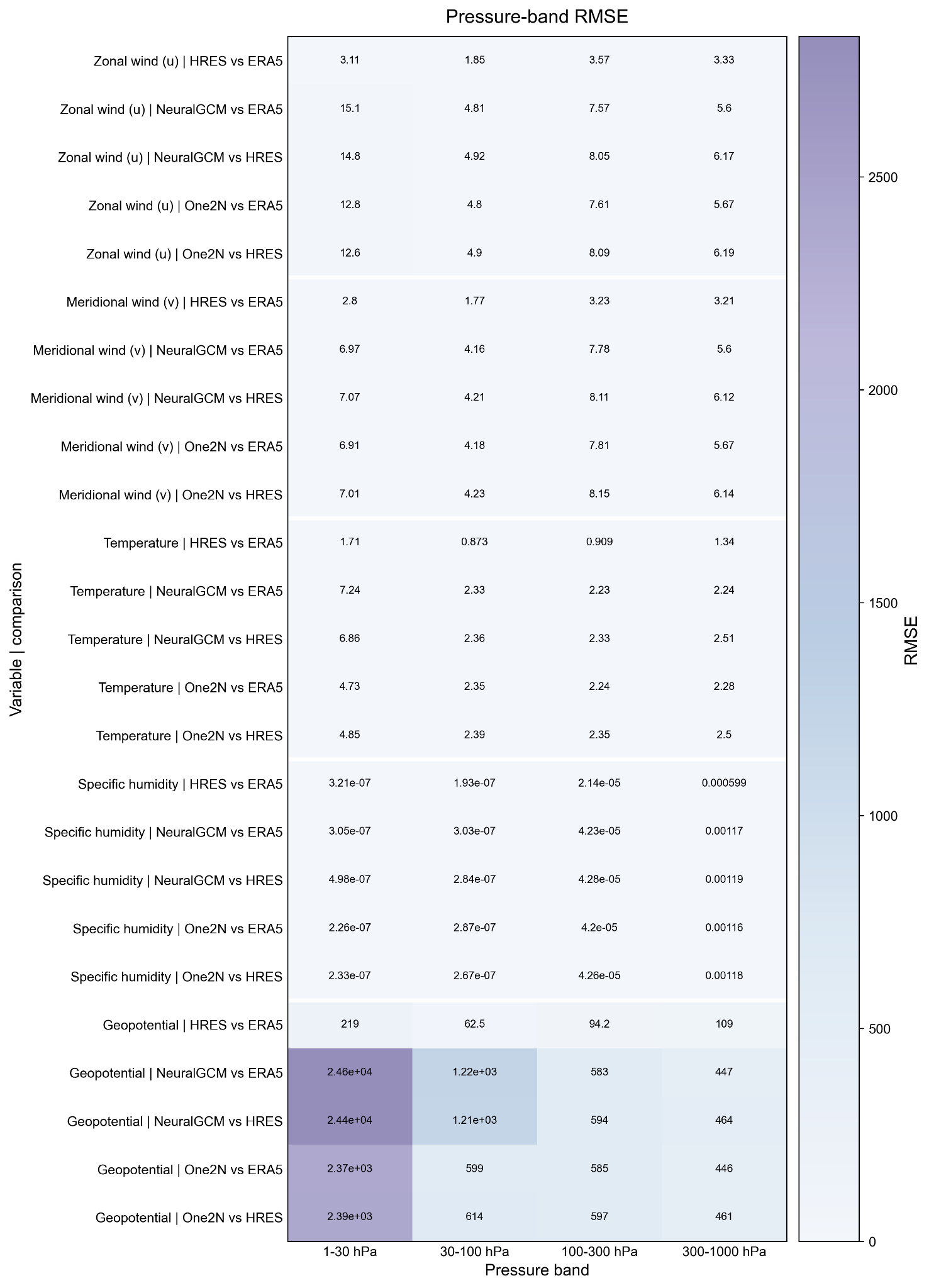}
\caption[Pressure-band RMSE against alternative reference analyses]{\textbf{Pressure-band RMSE against alternative reference analyses.}
As in Supplementary Fig.~\ref{fig:supp_pressure_nrmse}, but showing dimensional RMSE. Colours encode unscaled numerical values and should be interpreted only within each variable.}
\label{fig:supp_pressure_rmse}
\end{figure}

One2N-0.07\textdegree{} reduces upper-level RMSE against both references, with corresponding reductions in pressure-band nRMSE (Supplementary Figs.~\ref{fig:supp_pressure_nrmse} and~\ref{fig:supp_pressure_rmse}). In the 1--30~hPa band, geopotential nRMSE decreases from 2.600 to 0.273 against HRES-0.07\textdegree{} and from 2.614 to 0.271 against ERA5-0.25\textdegree{}, corresponding to reductions of 89.5\% and 89.6\%. Temperature nRMSE decreases by 28.4\% and 33.4\%, respectively. These improvements persist when verification changes from the HRES-0.07\textdegree{} analysis used to train the residual module to the alternative ERA5-0.25\textdegree{} reference.

The vertical dependence also persists across references. Geopotential nRMSE in the 30--100~hPa band decreases by 46.9\% against HRES-0.07\textdegree{} and 48.5\% against ERA5-0.25\textdegree{}, whereas differences between One2N-0.07\textdegree{} and NeuralGCM-0.7\textdegree{} remain small over most of the troposphere. Wind improvements are less uniform. Agreement between dimensional RMSE and nRMSE indicates that the broad vertical contrast is not solely a consequence of normalization. Upper-level humidity nRMSE is considered alongside dimensional errors and spatial correlation because the small reference variance can amplify normalized errors.

\subsection{Implications for the vertical contrast}
\label{subsec:vertical_interpretation}

The diagnostics connect the parent model's error structure with the distribution of One2N-0.07\textdegree{} gains. For the representative tropospheric metrics, small parent-model mean biases accompany broadly comparable One2N-0.07\textdegree{} forecast skill. In upper-stratospheric geopotential and temperature, large biases of consistent sign against both analyses coincide with the largest reductions in forecast error. This alignment supports a contribution from systematic bias correction to the upper-level improvements, alongside the spatial refinement illustrated in the main text.

The cross-reference comparison strengthens this interpretation by showing that the error reductions extend beyond the analysis used for residual training. It does not quantify the fraction of improvement attributable to bias correction or identify the physical source of the parent-model bias.

\section{Tropical cyclone tracking and verification}
\label{sec:supp_note7}

\subsection{Tracking procedure}
\label{subsec:tc_tracking}

Cyclone centres were tracked every 6~h, starting from the IBTrACS position at forecast initialization. Subsequent searches combined motion persistence with environmental steering estimated from winds at 850, 700, 500 and 200~hPa. The search radius ranged from 180 to 445~km according to forecast lead time and recent translation speed.

Candidates were identified from local minima of 1000-hPa geopotential in One2N-0.07\textdegree{} and NeuralGCM-0.7\textdegree{}, and from mean sea-level pressure minima in Pangu-Weather-0.25\textdegree{}. They were screened using three criteria:
\begin{enumerate}
\item Cyclonic 850-hPa relative vorticity exceeding \(5\times10^{-5}\ \mathrm{s^{-1}}\) in magnitude, with positive sign in the Northern Hemisphere and negative sign in the Southern Hemisphere.
\item Maximum 850-hPa wind speed exceeding \(8\ \mathrm{m\,s^{-1}}\).
\item A warm-core diagnostic based on the difference in \(Z_{200}-Z_{850}\) between the inner region (\(r<250\)~km) and surrounding annulus (\(250<r<500\)~km), evaluated against a lead-dependent threshold.
\end{enumerate}
Here, \(Z_p\) is geopotential at pressure \(p\) and \(r\) is distance from the candidate. The vorticity and wind criteria used local extrema within a diagnostic radius ranging from 120 to 278~km.

Retained candidates were ranked using the low-level minimum, vorticity, wind and warm-core signals, with penalties for implausible displacement or directional changes. A confidence-weighted combination of the selected candidate and predicted position supplied the updated track position. Thus, the models shared the screening and tracking framework, with different low-level detection fields. Motion-extrapolated positions occasionally used in the case-study displays were excluded from the statistical evaluation.

\subsection{Track error and paired comparisons}
\label{subsec:track_error_metric}

For a forecast centre \((\phi_f,\lambda_f)\) and reference centre \((\phi_o,\lambda_o)\) at the same valid time, track error is the great-circle distance

\begin{equation}\label{eq:track_error}
E_{\mathrm{track}}=
2R_e\arcsin\!\sqrt{a}, \quad 
a=\sin^2\!\left(\frac{\phi_f-\phi_o}{2}\right)+
\cos\phi_f\cos\phi_o\sin^2\!\left(\frac{\lambda_f-\lambda_o}{2}\right),
\end{equation}
where angles are in radians and \(R_e\) is Earth's mean radius; distances are reported in kilometres.

Let \(\mathcal M_\tau\) be the common set of cyclone--initialization pairs verified at lead time \(\tau\), and let \(E^{(a)}_{s,m}(\tau)\) denote the error of model \(a\) for cyclone \(s\) and initialization \(m\). The median error is

\begin{equation}\label{eq:tc_median_error}
\widetilde E^{(a)}(\tau)=
\operatorname{median}_{(s,m)\in\mathcal M_\tau}
E^{(a)}_{s,m}(\tau).
\end{equation}
The percentage reduction relative to comparator \(b\) is
\(100[1-\widetilde E^{(\text{One2N-0.07\textdegree{}})}(\tau)/\widetilde E^{(b)}(\tau)]\).

Paired error differences are calculated before summarization:

\begin{equation}\label{eq:tc_paired_difference}
d^{(b)}_{s,m}(\tau)=
E^{(b)}_{s,m}(\tau)-E^{(\text{One2N-0.07\textdegree{}})}_{s,m}(\tau).
\end{equation}
Positive differences indicate lower error for One2N-0.07\textdegree{}. The paired comparison reports the median of these differences and the fraction satisfying \(d^{(b)}_{s,m}(\tau)>0\), using the same \(\mathcal M_\tau\) for both comparators. The median paired difference is distinct from the difference between the two models' median errors.

\subsection{Evaluation samples and confidence intervals}
\label{subsec:tc_evaluation_protocol}

The evaluation used IBTrACS v04r01 best-track positions as the observational reference. Haikui, Saola and Bolaven provided the three case studies shown in the main text. The 2023 evaluation comprised 185 cyclone--initialization pairs from 84 cyclones. At each lead time, verification retained only matched samples with valid IBTrACS positions and detected candidates in all three models.

At each lead time, \(\mathcal M_\tau\) included only exact matches in cyclone identity, initialization time and valid time, with a valid best-track position and an objectively identified centre in all three models. Extrapolated positions were excluded. The number of matched forecasts, \(n_\tau=|\mathcal M_\tau|\), therefore varied with lead time. Initial positions at 0~h were prescribed from IBTrACS and excluded from forecast-error statistics. These comparisons quantify position error conditional on successful detection in all three models.

The pointwise 95\% confidence intervals shown for the median-error curves were estimated separately at each lead time from 2,000 bootstrap resamples with tropical cyclones as the resampling clusters. Each resample retained the matched forecast initializations belonging to the sampled cyclones, including their multiplicity when a cyclone was sampled more than once, and recomputed the median error at each lead. Resampling at cyclone level retains the dependence among forecasts of the same storm. The interval bounds are the 2.5th and 97.5th percentiles of the bootstrap medians.

\clearpage
\renewcommand{\refname}{Supplementary references}
\bibliographystyle{sn-nature}
\bibliography{supplementary-references}

\begin{thebibliography}{10}
\expandafter\ifx\csname url\endcsname\relax
  \def\url#1{\burl{#1}}\fi
\expandafter\ifx\csname urlprefix\endcsname\relax\def\urlprefix{URL }\fi
\providecommand{\bibinfo}[2]{#2}
\providecommand{\eprint}[2][]{\url{#2}}
\providecommand{\doi}[1]{\url{https://doi.org/#1}}
\bibcommenthead

\bibitem{prein2015review}
\bibinfo{author}{Prein, A.~F.} \emph{et~al.}
\newblock \bibinfo{title}{A review on regional convection-permitting climate
  modeling: demonstrations, prospects, and challenges}.
\newblock \emph{\bibinfo{journal}{Rev. Geophys.}}
  \textbf{\bibinfo{volume}{53}}, \bibinfo{pages}{323--361}
  (\bibinfo{year}{2015}).

\bibitem{satoh2019global}
\bibinfo{author}{Satoh, M.} \emph{et~al.}
\newblock \bibinfo{title}{Global cloud-resolving models}.
\newblock \emph{\bibinfo{journal}{Curr. Clim. Change Rep.}}
  \textbf{\bibinfo{volume}{5}}, \bibinfo{pages}{172--184}
  (\bibinfo{year}{2019}).

\bibitem{schulthess2019reflecting}
\bibinfo{author}{Schulthess, T.~C.} \emph{et~al.}
\newblock \bibinfo{title}{Reflecting on the goal and baseline for exascale
  computing: a roadmap based on weather and climate simulations}.
\newblock \emph{\bibinfo{journal}{Comput. Sci. Eng.}}
  \textbf{\bibinfo{volume}{21}}, \bibinfo{pages}{30--41}
  (\bibinfo{year}{2019}).

\bibitem{wedi2020baseline}
\bibinfo{author}{Wedi, N.~P.} \emph{et~al.}
\newblock \bibinfo{title}{A baseline for global weather and climate simulations
  at 1 km resolution}.
\newblock \emph{\bibinfo{journal}{J. Adv. Model. Earth Syst.}}
  \textbf{\bibinfo{volume}{12}}, \bibinfo{pages}{e2020MS002192}
  (\bibinfo{year}{2020}).

\bibitem{bauer2015quiet}
\bibinfo{author}{Bauer, P.}, \bibinfo{author}{Thorpe, A.} \&
  \bibinfo{author}{Brunet, G.}
\newblock \bibinfo{title}{The quiet revolution of numerical weather
  prediction}.
\newblock \emph{\bibinfo{journal}{Nature}} \textbf{\bibinfo{volume}{525}},
  \bibinfo{pages}{47--55} (\bibinfo{year}{2015}).

\bibitem{stevens2019dyamond}
\bibinfo{author}{Stevens, B.} \emph{et~al.}
\newblock \bibinfo{title}{{DYAMOND}: the dynamics of the atmospheric general
  circulation modeled on non-hydrostatic domains}.
\newblock \emph{\bibinfo{journal}{Prog. Earth Planet. Sci.}}
  \textbf{\bibinfo{volume}{6}}, \bibinfo{pages}{61} (\bibinfo{year}{2019}).

\bibitem{schar2020kilometer}
\bibinfo{author}{Sch{\"{a}}r, C.} \emph{et~al.}
\newblock \bibinfo{title}{Kilometer-scale climate models: prospects and
  challenges}.
\newblock \emph{\bibinfo{journal}{Bull. Am. Meteorol. Soc.}}
  \textbf{\bibinfo{volume}{101}}, \bibinfo{pages}{E567--E587}
  (\bibinfo{year}{2020}).

\bibitem{dueben2018challenges}
\bibinfo{author}{Dueben, P.~D.} \& \bibinfo{author}{Bauer, P.}
\newblock \bibinfo{title}{Challenges and design choices for global weather and
  climate models based on machine learning}.
\newblock \emph{\bibinfo{journal}{Geosci. Model Dev.}}
  \textbf{\bibinfo{volume}{11}}, \bibinfo{pages}{3999--4009}
  (\bibinfo{year}{2018}).

\bibitem{heinzeller2016towards}
\bibinfo{author}{Heinzeller, D.}, \bibinfo{author}{Duda, M.~G.} \&
  \bibinfo{author}{Kunstmann, H.}
\newblock \bibinfo{title}{Towards convection-resolving, global atmospheric
  simulations with the model for prediction across scales ({MPAS}) v3.1: an
  extreme scaling experiment}.
\newblock \emph{\bibinfo{journal}{Geosci. Model Dev.}}
  \textbf{\bibinfo{volume}{9}}, \bibinfo{pages}{77--110}
  (\bibinfo{year}{2016}).

\bibitem{klocke2025computing}
\bibinfo{author}{Klocke, D.} \emph{et~al.}
\newblock \bibinfo{title}{Computing the full {Earth} system at 1km resolution}.
\newblock In \emph{\bibinfo{booktitle}{Proceedings of the International
  Conference for High Performance Computing, Networking, Storage and
  Analysis}}, \bibinfo{pages}{125--136} (\bibinfo{publisher}{ACM},
  \bibinfo{year}{2025}).

\bibitem{bi2023accurate}
\bibinfo{author}{Bi, K.} \emph{et~al.}
\newblock \bibinfo{title}{Accurate medium-range global weather forecasting with
  {3D} neural networks}.
\newblock \emph{\bibinfo{journal}{Nature}} \textbf{\bibinfo{volume}{619}},
  \bibinfo{pages}{533--538} (\bibinfo{year}{2023}).

\bibitem{lam2023learning}
\bibinfo{author}{Lam, R.} \emph{et~al.}
\newblock \bibinfo{title}{Learning skillful medium-range global weather
  forecasting}.
\newblock \emph{\bibinfo{journal}{Science}} \textbf{\bibinfo{volume}{382}},
  \bibinfo{pages}{1416--1421} (\bibinfo{year}{2023}).

\bibitem{chen2023fuxi}
\bibinfo{author}{Chen, L.} \emph{et~al.}
\newblock \bibinfo{title}{{FuXi}: a cascade machine learning forecasting system
  for 15-day global weather forecast}.
\newblock \emph{\bibinfo{journal}{npj Clim. Atmos. Sci.}}
  \textbf{\bibinfo{volume}{6}}, \bibinfo{pages}{190} (\bibinfo{year}{2023}).

\bibitem{bodnar2025foundation}
\bibinfo{author}{Bodnar, C.} \emph{et~al.}
\newblock \bibinfo{title}{A foundation model for the {Earth} system}.
\newblock \emph{\bibinfo{journal}{Nature}} \textbf{\bibinfo{volume}{641}},
  \bibinfo{pages}{1180--1187} (\bibinfo{year}{2025}).

\bibitem{han2024fengwu}
\bibinfo{author}{Han, T.} \emph{et~al.}
\newblock \bibinfo{title}{{FengWu-GHR}: learning the kilometer-scale
  medium-range global weather forecasting}.
\newblock \bibinfo{howpublished}{Preprint at
  \url{https://arxiv.org/abs/2402.00059}} (\bibinfo{year}{2024}).

\bibitem{chen2016sublinear}
\bibinfo{author}{Chen, T.}, \bibinfo{author}{Xu, B.}, \bibinfo{author}{Zhang,
  C.} \& \bibinfo{author}{Guestrin, C.}
\newblock \bibinfo{title}{Training deep nets with sublinear memory cost}.
\newblock \bibinfo{howpublished}{Preprint at
  \url{https://arxiv.org/abs/1604.06174}} (\bibinfo{year}{2016}).

\bibitem{wattmeyer2021correcting}
\bibinfo{author}{Watt-Meyer, O.} \emph{et~al.}
\newblock \bibinfo{title}{Correcting weather and climate models by machine
  learning nudged historical simulations}.
\newblock \emph{\bibinfo{journal}{Geophys. Res. Lett.}}
  \textbf{\bibinfo{volume}{48}}, \bibinfo{pages}{e2021GL092555}
  (\bibinfo{year}{2021}).

\bibitem{kochkov2021accelerated}
\bibinfo{author}{Kochkov, D.} \emph{et~al.}
\newblock \bibinfo{title}{Machine learning--accelerated computational fluid
  dynamics}.
\newblock \emph{\bibinfo{journal}{Proc. Natl Acad. Sci. USA}}
  \textbf{\bibinfo{volume}{118}}, \bibinfo{pages}{e2101784118}
  (\bibinfo{year}{2021}).

\bibitem{pathak2020augment}
\bibinfo{author}{Pathak, J.} \emph{et~al.}
\newblock \bibinfo{title}{Using machine learning to augment coarse-grid
  computational fluid dynamics simulations}.
\newblock \bibinfo{howpublished}{Preprint at
  \url{https://arxiv.org/abs/2010.00072}} (\bibinfo{year}{2020}).

\bibitem{farchi2021correcting}
\bibinfo{author}{Farchi, A.}, \bibinfo{author}{Laloyaux, P.},
  \bibinfo{author}{Bonavita, M.} \& \bibinfo{author}{Bocquet, M.}
\newblock \bibinfo{title}{Using machine learning to correct model error in data
  assimilation and forecast applications}.
\newblock \emph{\bibinfo{journal}{Q. J. R. Meteorol. Soc.}}
  \textbf{\bibinfo{volume}{147}}, \bibinfo{pages}{3067--3084}
  (\bibinfo{year}{2021}).

\bibitem{kochkov2024neural}
\bibinfo{author}{Kochkov, D.} \emph{et~al.}
\newblock \bibinfo{title}{Neural general circulation models for weather and
  climate}.
\newblock \emph{\bibinfo{journal}{Nature}} \textbf{\bibinfo{volume}{632}},
  \bibinfo{pages}{1060--1066} (\bibinfo{year}{2024}).

\bibitem{hu2026hybrid}
\bibinfo{author}{Hu, Y.} \emph{et~al.}
\newblock \bibinfo{title}{A hybrid framework for global weather forecasting via
  low-resolution dynamical core and multigrid neural operator}.
\newblock \emph{\bibinfo{journal}{npj Clim. Atmos. Sci.}}
  \textbf{\bibinfo{volume}{9}}, \bibinfo{pages}{112} (\bibinfo{year}{2026}).

\bibitem{brenowitz2018prognostic}
\bibinfo{author}{Brenowitz, N.~D.} \& \bibinfo{author}{Bretherton, C.~S.}
\newblock \bibinfo{title}{Prognostic validation of a neural network unified
  physics parameterization}.
\newblock \emph{\bibinfo{journal}{Geophys. Res. Lett.}}
  \textbf{\bibinfo{volume}{45}}, \bibinfo{pages}{6289--6298}
  (\bibinfo{year}{2018}).

\bibitem{rasp2020coupled}
\bibinfo{author}{Rasp, S.}
\newblock \bibinfo{title}{Coupled online learning as a way to tackle
  instabilities and biases in neural network parameterizations: general
  algorithms and {Lorenz} 96 case study (v1.0)}.
\newblock \emph{\bibinfo{journal}{Geosci. Model Dev.}}
  \textbf{\bibinfo{volume}{13}}, \bibinfo{pages}{2185--2196}
  (\bibinfo{year}{2020}).

\bibitem{gilleland2009intercomparison}
\bibinfo{author}{Gilleland, E.}, \bibinfo{author}{Ahijevych, D.},
  \bibinfo{author}{Brown, B.~G.}, \bibinfo{author}{Casati, B.} \&
  \bibinfo{author}{Ebert, E.~E.}
\newblock \bibinfo{title}{Intercomparison of spatial forecast verification
  methods}.
\newblock \emph{\bibinfo{journal}{Weather Forecast.}}
  \textbf{\bibinfo{volume}{24}}, \bibinfo{pages}{1416--1430}
  (\bibinfo{year}{2009}).

\bibitem{pmlr-v267-subich25a}
\bibinfo{author}{Subich, C.}, \bibinfo{author}{Husain, S.~Z.},
  \bibinfo{author}{Separovic, L.} \& \bibinfo{author}{Yang, J.}
\newblock \bibinfo{title}{Fixing the double penalty in data-driven weather
  forecasting through a modified spherical harmonic loss function}.
\newblock In \emph{\bibinfo{booktitle}{Proceedings of the 42nd International
  Conference on Machine Learning}}, \bibinfo{editor}{Singh, A.} \emph{et~al.}
  (eds), Vol. \bibinfo{volume}{267} of \emph{\bibinfo{series}{Proc. Mach.
  Learn. Res.}}, \bibinfo{pages}{57191--57211} (\bibinfo{publisher}{PMLR},
  \bibinfo{year}{2025}).
\newblock \urlprefix\url{https://proceedings.mlr.press/v267/subich25a.html}.

\bibitem{cao2023swinunet}
\bibinfo{author}{Cao, H.} \emph{et~al.}
\newblock \bibinfo{title}{{Swin-Unet}: unet-like pure transformer for medical
  image segmentation}.
\newblock In \emph{\bibinfo{booktitle}{Computer Vision -- ECCV 2022
  Workshops}}, \bibinfo{editor}{Karlinsky, L.}, \bibinfo{editor}{Michaeli, T.}
  \& \bibinfo{editor}{Nishino, K.} (eds), Vol. \bibinfo{volume}{13803} of
  \emph{\bibinfo{series}{Lecture Notes in Computer Science}},
  \bibinfo{pages}{205--218} (\bibinfo{publisher}{Springer},
  \bibinfo{address}{Cham}, \bibinfo{year}{2023}).

\bibitem{liu2021swin}
\bibinfo{author}{Liu, Z.} \emph{et~al.}
\newblock \bibinfo{title}{{Swin Transformer}: hierarchical vision transformer
  using shifted windows}.
\newblock In \emph{\bibinfo{booktitle}{Proceedings of the IEEE/CVF
  International Conference on Computer Vision}}, \bibinfo{pages}{10012--10022}
  (\bibinfo{publisher}{IEEE}, \bibinfo{year}{2021}).

\bibitem{ecmwf2016hresdata}
\bibinfo{author}{{European Centre for Medium-Range Weather Forecasts}}.
\newblock \bibinfo{title}{{ECMWF IFS} high-resolution operational forecasts}.
\newblock \bibinfo{howpublished}{NSF National Center for Atmospheric Research}
  (\bibinfo{year}{2016}).
\newblock \urlprefix\url{https://doi.org/10.5065/D68050ZV}.

\bibitem{hersbach2020era5}
\bibinfo{author}{Hersbach, H.} \emph{et~al.}
\newblock \bibinfo{title}{The {ERA5} global reanalysis}.
\newblock \emph{\bibinfo{journal}{Q. J. R. Meteorol. Soc.}}
  \textbf{\bibinfo{volume}{146}}, \bibinfo{pages}{1999--2049}
  (\bibinfo{year}{2020}).

\bibitem{rasp2024weatherbench2}
\bibinfo{author}{Rasp, S.} \emph{et~al.}
\newblock \bibinfo{title}{{WeatherBench 2}: a benchmark for the next generation
  of data-driven global weather models}.
\newblock \emph{\bibinfo{journal}{J. Adv. Model. Earth Syst.}}
  \textbf{\bibinfo{volume}{16}}, \bibinfo{pages}{e2023MS004019}
  (\bibinfo{year}{2024}).

\bibitem{knapp2010ibtracs}
\bibinfo{author}{Knapp, K.~R.}, \bibinfo{author}{Kruk, M.~C.},
  \bibinfo{author}{Levinson, D.~H.}, \bibinfo{author}{Diamond, H.~J.} \&
  \bibinfo{author}{Neumann, C.~J.}
\newblock \bibinfo{title}{The international best track archive for climate
  stewardship ({IBTrACS}): unifying tropical cyclone data}.
\newblock \emph{\bibinfo{journal}{Bull. Am. Meteorol. Soc.}}
  \textbf{\bibinfo{volume}{91}}, \bibinfo{pages}{363--376}
  (\bibinfo{year}{2010}).

\bibitem{gahtan2024ibtracsdata}
\bibinfo{author}{Gahtan, J.} \emph{et~al.}
\newblock \bibinfo{title}{International best track archive for climate
  stewardship ({IBTrACS}) project, version 4r01}.
\newblock \bibinfo{howpublished}{NOAA National Centers for Environmental
  Information} (\bibinfo{year}{2024}).
\newblock \urlprefix\url{https://doi.org/10.25921/82ty-9e16}.

\bibitem{bonavita2016evolution}
\bibinfo{author}{Bonavita, M.}, \bibinfo{author}{H{\'{o}}lm, E.},
  \bibinfo{author}{Isaksen, L.} \& \bibinfo{author}{Fisher, M.}
\newblock \bibinfo{title}{The evolution of the {ECMWF} hybrid data assimilation
  system}.
\newblock \emph{\bibinfo{journal}{Q. J. R. Meteorol. Soc.}}
  \textbf{\bibinfo{volume}{142}}, \bibinfo{pages}{287--303}
  (\bibinfo{year}{2016}).

\bibitem{skamarock2012multiscale}
\bibinfo{author}{Skamarock, W.~C.} \emph{et~al.}
\newblock \bibinfo{title}{A multiscale nonhydrostatic atmospheric model using
  centroidal {Voronoi} tesselations and {C}-grid staggering}.
\newblock \emph{\bibinfo{journal}{Mon. Weather Rev.}}
  \textbf{\bibinfo{volume}{140}}, \bibinfo{pages}{3090--3105}
  (\bibinfo{year}{2012}).

\bibitem{zangl2015icon}
\bibinfo{author}{Z{\"{a}}ngl, G.}, \bibinfo{author}{Reinert, D.},
  \bibinfo{author}{R{\'{i}}podas, P.} \& \bibinfo{author}{Baldauf, M.}
\newblock \bibinfo{title}{The {ICON} (icosahedral non-hydrostatic) modelling
  framework of {DWD} and {MPI-M}: description of the non-hydrostatic dynamical
  core}.
\newblock \emph{\bibinfo{journal}{Q. J. R. Meteorol. Soc.}}
  \textbf{\bibinfo{volume}{141}}, \bibinfo{pages}{563--579}
  (\bibinfo{year}{2015}).

\bibitem{hess2025fast}
\bibinfo{author}{Hess, P.}, \bibinfo{author}{Aich, M.}, \bibinfo{author}{Pan,
  B.} \& \bibinfo{author}{Boers, N.}
\newblock \bibinfo{title}{Fast, scale-adaptive and uncertainty-aware
  downscaling of {Earth} system model fields with generative machine learning}.
\newblock \emph{\bibinfo{journal}{Nat. Mach. Intell.}}
  \textbf{\bibinfo{volume}{7}}, \bibinfo{pages}{363--373}
  (\bibinfo{year}{2025}).

\bibitem{mardani2025residual}
\bibinfo{author}{Mardani, M.} \emph{et~al.}
\newblock \bibinfo{title}{Residual corrective diffusion modeling for km-scale
  atmospheric downscaling}.
\newblock \emph{\bibinfo{journal}{Commun. Earth Environ.}}
  \textbf{\bibinfo{volume}{6}}, \bibinfo{pages}{124} (\bibinfo{year}{2025}).

\bibitem{brenowitz2020interpreting}
\bibinfo{author}{Brenowitz, N.~D.}, \bibinfo{author}{Beucler, T.},
  \bibinfo{author}{Pritchard, M.} \& \bibinfo{author}{Bretherton, C.~S.}
\newblock \bibinfo{title}{Interpreting and stabilizing machine-learning
  parametrizations of convection}.
\newblock \emph{\bibinfo{journal}{J. Atmos. Sci.}}
  \textbf{\bibinfo{volume}{77}}, \bibinfo{pages}{4357--4375}
  (\bibinfo{year}{2020}).

\bibitem{zhang2022integral}
\bibinfo{author}{Zhang, H.} \& \bibinfo{author}{Tian, X.}
\newblock \bibinfo{title}{Integral correction of initial and model errors in
  system of multigrid {NLS-4DVar} data assimilation for numerical weather
  prediction ({SNAP})}.
\newblock \emph{\bibinfo{journal}{Q. J. R. Meteorol. Soc.}}
  \textbf{\bibinfo{volume}{148}}, \bibinfo{pages}{2490--2506}
  (\bibinfo{year}{2022}).

\bibitem{he2016deep}
\bibinfo{author}{He, K.}, \bibinfo{author}{Zhang, X.}, \bibinfo{author}{Ren,
  S.} \& \bibinfo{author}{Sun, J.}
\newblock \bibinfo{title}{Deep residual learning for image recognition}.
\newblock In \emph{\bibinfo{booktitle}{Proceedings of the IEEE Conference on
  Computer Vision and Pattern Recognition}}, \bibinfo{pages}{770--778}
  (\bibinfo{publisher}{IEEE}, \bibinfo{year}{2016}).

\bibitem{ronneberger2015unet}
\bibinfo{author}{Ronneberger, O.}, \bibinfo{author}{Fischer, P.} \&
  \bibinfo{author}{Brox, T.}
\newblock \bibinfo{title}{{U-Net}: convolutional networks for biomedical image
  segmentation}.
\newblock In \emph{\bibinfo{booktitle}{Medical Image Computing and
  Computer-Assisted Intervention -- MICCAI 2015}}, Vol. \bibinfo{volume}{9351}
  of \emph{\bibinfo{series}{Lecture Notes in Computer Science}},
  \bibinfo{pages}{234--241} (\bibinfo{publisher}{Springer},
  \bibinfo{address}{Cham}, \bibinfo{year}{2015}).

\bibitem{battaglia2018relational}
\bibinfo{author}{Battaglia, P.~W.} \emph{et~al.}
\newblock \bibinfo{title}{Relational inductive biases, deep learning, and graph
  networks}.
\newblock \bibinfo{howpublished}{Preprint at
  \url{https://arxiv.org/abs/1806.01261}} (\bibinfo{year}{2018}).

\bibitem{vaswani2017attention}
\bibinfo{author}{Vaswani, A.} \emph{et~al.}
\newblock \bibinfo{title}{Attention is all you need}.
\newblock In \emph{\bibinfo{booktitle}{Advances in Neural Information
  Processing Systems}}, Vol.~\bibinfo{volume}{30}, \bibinfo{pages}{5998--6008}
  (\bibinfo{publisher}{Curran Associates, Inc.}, \bibinfo{year}{2017}).

\bibitem{kashinath2021physics}
\bibinfo{author}{Kashinath, K.} \emph{et~al.}
\newblock \bibinfo{title}{Physics-informed machine learning: case studies for
  weather and climate modelling}.
\newblock \emph{\bibinfo{journal}{Phil. Trans. R. Soc. A}}
  \textbf{\bibinfo{volume}{379}}, \bibinfo{pages}{20200093}
  (\bibinfo{year}{2021}).

\bibitem{c3s2018era5data}
\bibinfo{author}{{Copernicus Climate Change Service}}.
\newblock \bibinfo{title}{{ERA5} hourly data on pressure levels from 1940 to
  present}.
\newblock \bibinfo{howpublished}{Copernicus Climate Change Service (C3S)
  Climate Data Store (CDS)} (\bibinfo{year}{2018}).
\newblock \urlprefix\url{https://doi.org/10.24381/cds.bd0915c6}.

\end{thebibliography}


\begin{thebibliography}{10}
\expandafter\ifx\csname url\endcsname\relax
  \def\url#1{\burl{#1}}\fi
\expandafter\ifx\csname urlprefix\endcsname\relax\def\urlprefix{URL }\fi
\providecommand{\bibinfo}[2]{#2}
\providecommand{\eprint}[2][]{\url{#2}}
\providecommand{\doi}[1]{\url{https://doi.org/#1}}
\bibcommenthead

\bibitem{ecmwf2016hresdata}
\bibinfo{author}{{European Centre for Medium-Range Weather Forecasts}}.
\newblock \bibinfo{title}{{ECMWF IFS} high-resolution operational forecasts}.
\newblock \bibinfo{howpublished}{NSF National Center for Atmospheric Research}
  (\bibinfo{year}{2016}).
\newblock \urlprefix\url{https://doi.org/10.5065/D68050ZV}.

\bibitem{hersbach2020era5}
\bibinfo{author}{Hersbach, H.} \emph{et~al.}
\newblock \bibinfo{title}{The {ERA5} global reanalysis}.
\newblock \emph{\bibinfo{journal}{Q. J. R. Meteorol. Soc.}}
  \textbf{\bibinfo{volume}{146}}, \bibinfo{pages}{1999--2049}
  (\bibinfo{year}{2020}).

\bibitem{c3s2018era5data}
\bibinfo{author}{{Copernicus Climate Change Service}}.
\newblock \bibinfo{title}{{ERA5} hourly data on pressure levels from 1940 to
  present}.
\newblock \bibinfo{howpublished}{Copernicus Climate Change Service (C3S)
  Climate Data Store (CDS)} (\bibinfo{year}{2018}).
\newblock \urlprefix\url{https://doi.org/10.24381/cds.bd0915c6}.

\bibitem{knapp2010ibtracs}
\bibinfo{author}{Knapp, K.~R.}, \bibinfo{author}{Kruk, M.~C.},
  \bibinfo{author}{Levinson, D.~H.}, \bibinfo{author}{Diamond, H.~J.} \&
  \bibinfo{author}{Neumann, C.~J.}
\newblock \bibinfo{title}{The international best track archive for climate
  stewardship ({IBTrACS}): unifying tropical cyclone data}.
\newblock \emph{\bibinfo{journal}{Bull. Am. Meteorol. Soc.}}
  \textbf{\bibinfo{volume}{91}}, \bibinfo{pages}{363--376}
  (\bibinfo{year}{2010}).

\bibitem{gahtan2024ibtracsdata}
\bibinfo{author}{Gahtan, J.} \emph{et~al.}
\newblock \bibinfo{title}{International best track archive for climate
  stewardship ({IBTrACS}) project, version 4r01}.
\newblock \bibinfo{howpublished}{NOAA National Centers for Environmental
  Information} (\bibinfo{year}{2024}).
\newblock \urlprefix\url{https://doi.org/10.25921/82ty-9e16}.

\bibitem{jones1999conservative}
\bibinfo{author}{Jones, P.~W.}
\newblock \bibinfo{title}{First- and second-order conservative remapping
  schemes for grids in spherical coordinates}.
\newblock \emph{\bibinfo{journal}{Mon. Weather Rev.}}
  \textbf{\bibinfo{volume}{127}}, \bibinfo{pages}{2204--2210}
  (\bibinfo{year}{1999}).

\bibitem{kochkov2024neural}
\bibinfo{author}{Kochkov, D.} \emph{et~al.}
\newblock \bibinfo{title}{Neural general circulation models for weather and
  climate}.
\newblock \emph{\bibinfo{journal}{Nature}} \textbf{\bibinfo{volume}{632}},
  \bibinfo{pages}{1060--1066} (\bibinfo{year}{2024}).

\bibitem{bi2023accurate}
\bibinfo{author}{Bi, K.} \emph{et~al.}
\newblock \bibinfo{title}{Accurate medium-range global weather forecasting with
  {3D} neural networks}.
\newblock \emph{\bibinfo{journal}{Nature}} \textbf{\bibinfo{volume}{619}},
  \bibinfo{pages}{533--538} (\bibinfo{year}{2023}).

\bibitem{cao2023swinunet}
\bibinfo{author}{Cao, H.} \emph{et~al.}
\newblock \bibinfo{title}{{Swin-Unet}: unet-like pure transformer for medical
  image segmentation}.
\newblock In \emph{\bibinfo{booktitle}{Computer Vision -- ECCV 2022
  Workshops}}, \bibinfo{editor}{Karlinsky, L.}, \bibinfo{editor}{Michaeli, T.}
  \& \bibinfo{editor}{Nishino, K.} (eds), Vol. \bibinfo{volume}{13803} of
  \emph{\bibinfo{series}{Lecture Notes in Computer Science}},
  \bibinfo{pages}{205--218} (\bibinfo{publisher}{Springer},
  \bibinfo{address}{Cham}, \bibinfo{year}{2023}).

\bibitem{liu2021swin}
\bibinfo{author}{Liu, Z.} \emph{et~al.}
\newblock \bibinfo{title}{{Swin Transformer}: hierarchical vision transformer
  using shifted windows}.
\newblock In \emph{\bibinfo{booktitle}{Proceedings of the IEEE/CVF
  International Conference on Computer Vision}}, \bibinfo{pages}{10012--10022}
  (\bibinfo{publisher}{IEEE}, \bibinfo{year}{2021}).

\bibitem{lam2023learning}
\bibinfo{author}{Lam, R.} \emph{et~al.}
\newblock \bibinfo{title}{Learning skillful medium-range global weather
  forecasting}.
\newblock \emph{\bibinfo{journal}{Science}} \textbf{\bibinfo{volume}{382}},
  \bibinfo{pages}{1416--1421} (\bibinfo{year}{2023}).

\bibitem{chen2023fuxi}
\bibinfo{author}{Chen, L.} \emph{et~al.}
\newblock \bibinfo{title}{{FuXi}: a cascade machine learning forecasting system
  for 15-day global weather forecast}.
\newblock \emph{\bibinfo{journal}{npj Clim. Atmos. Sci.}}
  \textbf{\bibinfo{volume}{6}}, \bibinfo{pages}{190} (\bibinfo{year}{2023}).

\bibitem{bodnar2025foundation}
\bibinfo{author}{Bodnar, C.} \emph{et~al.}
\newblock \bibinfo{title}{A foundation model for the {Earth} system}.
\newblock \emph{\bibinfo{journal}{Nature}} \textbf{\bibinfo{volume}{641}},
  \bibinfo{pages}{1180--1187} (\bibinfo{year}{2025}).

\end{thebibliography}

\end{document}